\documentclass[twocolumn,showpacs,aps,prd,floatfix]{revtex4}
\usepackage{graphicx}
\usepackage{dcolumn}
\usepackage{multirow}
\usepackage{amsmath}
\usepackage{amssymb}
\usepackage{epsfig}
\usepackage{color}
\usepackage{verbatim} % needed for multi-line comments
\usepackage{rotating} % needed for sidewaystable
\usepackage{hhline}
\usepackage{tabularx}

\newcommand{\BABARPubYear}    {11}
\newcommand{\BABARPubNumber}  {014}
\newcommand{\SLACPubNumber} {14811}
\newcommand{\LANLNumber} {1112.0702}

\input babarsym

\newcommand {\Bxlnu}  {\ensuremath{\Bbar \rightarrow X \ell \bar{\nu}}\xspace}
\newcommand {\Bxclnu} {\ensuremath{\Bbar \rightarrow X_c \ell \bar{\nu}}\xspace}
\newcommand {\Bxulnu} {\ensuremath{\Bbar \rightarrow X_u \ell \bar{\nu}}\xspace}
\newcommand {\Bdlnu} {\ensuremath{\Bbar \rightarrow D \ell \bar{\nu}}\xspace}
\newcommand {\Bdslnu} {\ensuremath{\Bbar \rightarrow D^* \ell \bar{\nu}}\xspace}
\newcommand {\Bdsslnu} {\ensuremath{\Bbar \rightarrow D^{**} \ell \bar{\nu}}\xspace}
\newcommand {\Bdbslnu} {\ensuremath{\Bbar \rightarrow D^{(*)} \ell \bar{\nu}}\xspace}
\newcommand {\Bdbsxlnu} {\ensuremath{\Bbar \rightarrow D^{(*)} X \ell \bar{\nu}}\xspace}

\newcommand {\mx}     {\ensuremath{M_{X}}\xspace}
\newcommand {\Pplus}  {\ensuremath{P_{+}}\xspace}

\newcommand {\Pl}     {\ensuremath{p_{\ell}^*}\xspace}
\newcommand {\Q}      {\ensuremath{q^{2}}\xspace}

\newcommand {\mX}     {\ensuremath{M_{X}}\xspace}

\newcommand {\breco}  {\ensuremath{B_\mathrm{reco}}\xspace}
\newcommand {\brecoil}{\ensuremath{B_\mathrm{recoil}}\xspace}
\newcommand {\D}      {\ensuremath{D}\xspace}
\newcommand {\beq}    {\begin{equation}}
\newcommand {\beqa}   {\begin{eqnarray}}
\newcommand {\beqn}   {\begin{eqnarray}}
\newcommand {\eeq}    {\end{equation}}
\newcommand {\eeqa}   {\end{eqnarray}}
\newcommand {\eeqn}   {\end{eqnarray}}
\makeatletter
\def\slash#1{{\mathpalette\c@ncel{#1}}} % TeXbook, bottom of p360
\makeatother
\newcommand{\lonesf} {\ensuremath{\lambda_1^{SF}}}
\newcommand{\lbarsf} {\ensuremath{\bar{\Lambda}^{SF}}}

\newcommand{\gevccsq}{\ensuremath{{\mathrm{\,Ge\kern -0.1em V^2}}}\xspace} %ok, c=1

\long\def\inst#1{\par\nobreak\kern 4pt\nobreak
    {\it #1}\par\vskip 10pt plus 3pt minus 3pt}

\begin{document}
\begin{flushleft}
\babar-PUB-\BABARPubYear/\BABARPubNumber \\
SLAC-PUB-\SLACPubNumber\\
hep-ex/\LANLNumber\\[5mm]
\end{flushleft}

% Title of the paper
\title{\large \bf Study of $\Bxulnu$ decays in \BB\ events tagged by 
a fully reconstructed $B$-meson decay and determination of \Vub} 

%% author list as of 01-Jun-2011 (386 authors)
%
\author{J.~P.~Lees}
\author{V.~Poireau}
\author{V.~Tisserand}
\affiliation{Laboratoire d'Annecy-le-Vieux de Physique des Particules (LAPP), Universit\'e de Savoie, CNRS/IN2P3,  F-74941 Annecy-Le-Vieux, France}
\author{J.~Garra~Tico}
\author{E.~Grauges}
\affiliation{Universitat de Barcelona, Facultat de Fisica, Departament ECM, E-08028 Barcelona, Spain }
\author{M.~Martinelli$^{ab}$}
\author{D.~A.~Milanes$^{a}$}
\author{A.~Palano$^{ab}$ }
\author{M.~Pappagallo$^{ab}$ }
\affiliation{INFN Sezione di Bari$^{a}$; Dipartimento di Fisica, Universit\`a di Bari$^{b}$, I-70126 Bari, Italy }
\author{G.~Eigen}
\author{B.~Stugu}
\affiliation{University of Bergen, Institute of Physics, N-5007 Bergen, Norway }
% ...
\collaboration{The \babar\ Collaboration}
\noaffiliation

\begin{abstract}
We report measurements of partial branching fractions for inclusive charmless semileptonic $B$
decays $\kern 0.18em\overline{\kern -0.18em B}{} \rightarrow X_u \ell \bar{\nu}$, 
and the determination of the CKM matrix element \ensuremath{|V_{ub}|}.
The analysis is based on a sample of 467 million $\Upsilon{(4S)} \to B\kern 0.18em\overline{\kern -0.18em B}{}\xspace$
decays recorded with the \mbox{\slshape B\kern-0.1em{\smaller A}\kern-0.1em B\kern-0.1em{\smaller A\kern-0.2em R}} 
detector at the PEP-II $e^{+} e^{-}$ storage rings. 
We select events in which the decay of one of the $B$ mesons is fully reconstructed 
and an electron or a muon signals the semileptonic decay of the other $B$ meson.  
We measure partial branching fractions $\Delta {\cal{B}}$ in several restricted regions of phase 
space and determine the CKM element \ensuremath{|V_{ub}|} based on different QCD predictions.
For decays with a charged lepton momentum  $p_{\ell}^*>1.0$~\ensuremath{{\mathrm{\,Ge\kern -0.1em V}}} 
in the $B$ meson rest frame, 
we obtain $\Delta {\cal{B}} =(1.80 \pm 0.13_{\rm stat.} \pm 0.15_{\rm sys.} \pm 0.02_{\rm theo.}) \times 10^{-3}$ 
from a fit to the two-dimensional \ensuremath{M_{X}}\ -- \ensuremath{q^{2}} distribution.  
Here, \ensuremath{M_{X}} refers to the invariant mass of the final state hadron $X$
and \ensuremath{q^{2}} is the invariant mass squared of the charged lepton and neutrino. 
From this measurement we extract  $|V_{ub}| = (4.33\pm 0.24_{\rm exp.} \pm 0.15_{\rm theo.}) \times 10^{-3}$ 
as the arithmetic average of four results obtained from four different QCD predictions of the partial rate.  
We separately determine partial branching fractions for \ensuremath{\kern 0.18em\overline{\kern -0.18em B}{}^0} and \ensuremath{B^-} 
decays and derive a limit on the isospin breaking in \ensuremath{\Bbar \rightarrow X_u \ell \bar{\nu}} decays.

\end{abstract}

                 % properties of bottom mesons

\maketitle  

\section{Introduction} 
A principal physics goal of the \babar\ experiment is to establish
\CP violation in \B meson decays and to test whether the observed effects
are consistent with the Standard Model (SM) expectations. In the SM,
\CP-violating effects result  from an irreducible phase in the 
Cabibbo-Kobayashi-Maskawa (CKM) quark-mixing matrix~\cite{Cabibbo:1963yz,Kobayashi:1973fv}.
Precise determinations of the magnitude of the matrix element \Vub will permit more stringent 
tests of the SM mechanism for \CP violation. This is best illustrated in terms of the 
unitarity triangle~\cite{Bjorken:1990rr}, the graphical representation of one of the unitarity conditions 
of the CKM matrix, for which the side opposite to the angle $\beta$ is proportional to the 
ratio $\Vub/|V_{cb}|$.
The best way to determine \Vub is to measure the decay rate for
\Bxulnu (here $X$ refers to a hadronic final state and the 
index $c$ or $u$ indicates whether this state carries charm or not), which is proportional to 
$\Vub^2$. 

There are two approaches to these measurements, based on either inclusive or exclusive 
measurements of semileptonic decays. The experimental uncertainties on the methods are largely independent, 
and the extraction of \Vub from the measured branching fractions relies on different 
sets of calculations of the hadronic contributions to the matrix element.  
For quite some time, the results of  measurements of \Vub\ from inclusive and exclusive decays 
have been only marginally consistent~\cite{kowalewski_mannel,CKMBook}.
Global fits~\cite{utfit,ckmfitter} testing the compatibility of the measured 
angles and sides with the unitarity triangle of the CKM matrix reveal small 
differences that might indicate potential deviations from SM expectations.  
Therefore, it is important to perform redundant and improved measurements, employing
different experimental techniques and a variety of theoretical
calculations, to better assess the accuracy of the theoretical and experimental uncertainties. 

Although inclusive branching fractions exceed those of individual exclusive decays by an order of 
magnitude, the most challenging task for inclusive measurements is the discrimination between 
the rare charmless signal and the much more abundant decays involving charmed mesons. 
To improve the signal-to-background ratio, the events are restricted to selected regions of phase space. 
Unfortunately these restrictions lead to difficulties in calculating partial branching fractions.
They impact the convergence of Heavy Quark Expansions (HQE)~\cite{Manohar:1993qn,Bigi:1993fm}, 
enhance perturbative and  
nonperturbative QCD corrections, and thus lead to significantly larger theoretical uncertainties 
in the determination of \Vub.

We report herein measurements of partial branching fractions $(\Delta {\cal{B}})$ 
for inclusive charmless semileptonic \B meson decays, \Bxulnu~\cite{footnote}.
This analysis 
extends the event selection and methods employed previously by \babar\ to a larger 
dataset~\cite{babar_inclusive}.   
We tag $\FourS \to \BB$ events with a fully reconstructed hadronic decay 
of one of the \B\ mesons (\breco). This technique results  
in a low event selection efficiency, but it uniquely determines the momentum and charge of 
both \B\ mesons in the event, reducing backgrounds significantly.
For charged $B$ mesons it also determines their flavor.
The semileptonic decay of the second \B\ meson 
(\brecoil) is identified by the presence of an electron or a muon and its kinematics are 
constrained such that the undetectable neutrino can be identified from the missing momentum 
and energy of the rest of the event. 
However, undetected and poorly reconstructed 
charged particles or photons lead to large backgrounds from the
dominant \Bxclnu decays, and they distort the kinematics, ${\it e.g.}$, 
the hadronic mass \mX and the leptonic mass squared \Q.

For the \breco sample, the two dominant background sources are non-\BB events from continuum processes, 
$\epem \to \qqbar (\gamma)$ with $q = \u$, \d, \s, or \c, and combinatorial \BB\ background.  
The sum of these two backgrounds is estimated from the distribution 
of the beam energy-substituted mass \mes, which takes the following 
form in the laboratory frame:
$\mes = \sqrt{(s/2+\vec{p}_B\cdot\vec{p}_\mathrm{beams})^2/E^2_\mathrm{beams}-\vec{p}^{\, 2}_B}$.
Here $\vec{p}_B$ refers to the momentum of the \breco candidate  
derived from the measured momenta of its decay products, 
$P_\mathrm{beams} = (E_\mathrm{beams},\vec{p}_\mathrm{beams})$
to the four-momentum of the colliding beam particles, and $\sqrt{s}$ to the total energy in the
\FourS frame.
For correctly reconstructed \breco decays, the distribution peaks at the \B meson mass, 
and the width of the peak is determined by the energy spread of the
colliding beams. The size of the underlying background is determined from a 
fit to the \mes distribution. 

We minimize experimental systematic uncertainties,
by measuring the yield for selected charmless semileptonic decays relative 
to the total yield of semileptonic decays \Bxlnu, after subtracting
combinatorial backgrounds of the \breco selection from both samples.

In order to reduce the overall uncertainties, measurement of the signal \Bxulnu
decays is restricted to regions of phase space where the background from the dominant 
\Bxclnu decays is suppressed and theoretical uncertainties can be reliably assessed.
Specifically, signal events tend to have higher charged lepton momenta in the 
$B$-meson rest frame (\Pl), lower \mx, 
higher \Q, and smaller values of the light-cone 
momentum $P_+=E_{X}-|\vec{p}_X|$, where $E_{X}$ and $\vec{p}_X$ are energy and momentum of the 
hadronic system $X$ in the \B meson rest frame.  

The observation of charged leptons with momenta exceeding the kinematic limit for \Bxclnu 
presented first evidence for charmless semileptonic decays.
This was followed by a series of measurements close 
to this kinematic limit~\cite{Bartelt:1993xh, Bornheim:2002du,Limosani:2005pi, Aubert:2005mg,Aubert:2005im}.  
Although the signal-to-background ratio for this small region of phase space is favorable, 
the theoretical uncertainties are large and difficult to quantify.  
Since then, efforts have been made to select larger phase space regions, thereby reducing the 
theoretical uncertainties. The Belle Collaboration has recently
published an analysis that covers about 88\% of the signal phase space~\cite{Belle_multivariate},
similarly to one of the studies detailed in this article.
 
We extract \Vub from the partial branching fractions relying on four different 
QCD calculations of the partial decay rate in several phase space regions: 
BLNP by Bosch, Lange, Neubert, and Paz~\cite{Lange:2005yw,Bosch:2004th,Bosch:2004cb}; 
DGE, the dressed gluon exponentiation by 
Andersen and Gardi~\cite{Andersen:2005mj,Gardi:2008bb}; 
ADFR by Aglietti, Di~Lodovico, Ferrara, and Ricciardi~\cite{Aglietti,Aglietti:2006yb}; 
and GGOU by Gambino, Giordano, Ossola and Uraltsev~\cite{GGOU}.
These calculations differ significantly in their treatment of perturbative corrections and the 
parameterization of nonperturbative effects that become important for the different restrictions 
in phase space.

This measurement of \Vub is based on combined samples of charged and neutral $B$ mesons.
In addition, we present measurements of the partial decays rates for $\Bzb$ and $B^-$ decays separately.  
The observed rates are found to be equal within uncertainties.  
We use this observation to set a limit on weak annihilation (WA), the process 
$b\ubar \to \ell^- \bar{\nu}_{\ell}$,
which is not included in the QCD calculation of the \Bxlnu decay rates.  
Since final state hadrons originate from soft gluon emission, 
WA is expected to contribute to the decay rate at large values of 
\Q~\cite{Bigi:1993bh,Voloshin:2001xi,Ligeti:2010vd,GambinoWA}. 

The outline of this paper is as follows: a brief overview of the \babar\ detector, particle
reconstruction and the data 
and Monte Carlo (MC) samples is given in Section~\ref{sec:detector}, followed in 
Section~\ref{sec:selection} by a description of the event reconstruction and selection 
of the two event samples, the charmless semileptonic signal sample and the inclusive 
semileptonic sample that serves as normalization. 
The measurement of the partial branching fractions and their systematic uncertainties are 
presented in Sections~\ref{meastec} and~\ref{sec:systematics}.   
The extraction of \Vub\ based on four sets of QCD calculations for seven selected regions 
of phase space is presented in Section~\ref{sec:theonew}, followed by the conclusions in 
Section~\ref{sec:conclusions}.

\section{Data Sample, Detector,  and Simulation}
\label{sec:detector}

\subsection{Data sample}
The data used in this analysis were recorded with the \babar\ detector at the \pep2 
asymmetric energy $e^+e^-$ collider operating at the \FourS\ resonance.
The total data sample, corresponding to an integrated luminosity of 426~$\mathrm{fb}^{-1}$
and containing 467 million $\FourS \to \BB$ events, was analyzed.

\subsection{The \babar\ detector}
The \babar\ detector and the general event reconstruction are described in
detail elsewhere~\cite{babar_NIM,Aubert:2002rg}.
For this analysis, the most important detector features are the charged-particle tracking, photon 
reconstruction, and particle identification.
The momenta and angles of charged
particles are measured in a tracking system consisting of a five-layer silicon vertex tracker (SVT) 
and a 40-layer, small-cell drift chamber (DCH). 
Charged particles of different masses are distinguished by their
ionization energy loss in the tracking devices and by the DIRC, a ring-imaging 
detector of internally reflected Cherenkov radiation. 
A finely segmented electromagnetic calorimeter (EMC) consisting of 6580 CsI(Tl) crystals measures 
the energy and position of showers generated by electrons and photons. 
The EMC is surrounded by a thin superconducting solenoid providing a 1.5~T magnetic field and by a 
steel flux return with a hexagonal barrel section and
two endcaps. The segmented flux return (IFR) is instrumented with multiple layers of resistive plate 
chambers (RPC) and limited streamer tubes (LST) to identify muons and to a lesser degree $K_L$. 

\subsection{Single particle reconstruction}
In order to reject misidentified and background tracks that do not originate from the interaction point, 
we require the radial and longitudinal impact parameters to be 
$r_0 < 1.5$~\cm and $|z_0| < 10$~\cm. 
For secondary tracks from $K_S \to \pi^+ \pi^-$ decays, no restrictions on the impact parameter are imposed. 
The efficiency for the reconstruction of charged particles inside the
fiducial volume for SVT, DCH, and EMC, defined by the polar angle
in the laboratory frame, $0.410 < \theta_{\mathrm{lab}} < 2.54$~rad,
exceeds 96\% and is well reproduced by Monte Carlo (MC) simulation. 

Electromagnetic showers are detected in the EMC as clusters of energy depositions.  
Photons are required not to be matched to a charged track extrapolated to the 
position of the shower maximum in the EMC.
To suppress photons from beam-related background, we only retain
photons with energies larger than $50$~\mev.
Clusters created by neutral hadrons ($K_L$ or neutrons) interacting in the EMC
are distinguished from photons by their shower shape.

Electrons are primarily separated from charged hadrons on the basis of the 
ratio of the energy deposited in the EMC to the track momentum. 
This quantity should be close to 1 for electrons since they deposit 
all their energy in the calorimeter. Most other charged tracks are minimum ionizing,
unless they shower in the EMC crystals. 

Muons are identified by a neural network that combines information
from the IFR with the measured track momentum and the energy deposition in the EMC.

The average
electron efficiency for laboratory momenta above 0.5~\gevc is $93\%$, largely
independent of momentum.  The average hadron misidentification rate is less than $0.2\%$. 
Within the polar-angle acceptance, the average muon efficiency rises with 
laboratory momentum and reaches a plateau of about $70\%$ above 1.4~\gevc. 
The muon efficiency varies between $50\%$ and $80\%$ as a function of the polar angle. 
The average hadron misidentification rate is about $1.5\%$, varying by about $0.5\%$ as a function 
of momentum and polar angle.
 
Charged kaons are selected on the basis of information from the DIRC, DCH, and SVT. 
The efficiency is higher than $80\%$ 
over most of the momentum range and varies with the polar angle. 
The probability of a pion to be misidentified as a kaon is close to $2\%$,
varying by about $1\%$ as a function of momentum and polar angle.

Neutral pions are reconstructed from pairs of photon candidates that are
detected in the EMC and are assumed to originate from
the primary vertex. Photon pairs having an invariant mass within 17.5~\mevcc 
(corresponding to $2.5~\sigma$) of the nominal \piz\ mass are considered \piz\ candidates. 
The overall detection efficiency, including solid angle restrictions, 
varies between 55\% and 65\% for \piz\ energies in the range of 0.2 to 2.5~\gev.

$\KS \to \pi^+ \pi^-$ decays are reconstructed as pairs of tracks of opposite charge with
a common vertex displaced from the interaction point. The invariant mass of the pair is required 
to be in the range $490 < m_{\pi^+\pi^-} < 505$~\mevcc. 

\subsection{Monte Carlo simulation}
\label{sec:signalgene}
We use MC techniques to simulate the response of the \babar\ 
detector~\cite{Agostinelli:2002hh} and the particle production and decays~\cite{evtgen}, 
to optimize selection criteria and to determine signal efficiencies and background distributions.
The agreement of the simulated distributions with those in data has been verified with control samples, 
as shown in Section~\ref{sec:thecomparison};
the impact of the inaccuracies of the simulation is estimated in Section~\ref{sec:systematics}.

The size of the simulated sample of generic \BB\ events exceeds the \BB\ data sample by about a 
factor of three.  This sample includes the common $\Bxclnu$ decays.
MC samples for inclusive and exclusive $\Bxulnu$ decays exceed the size of the
data samples by factors of 15 or more.

Charmless semileptonic \Bxulnu\ decays  are simulated as a combination of resonant three-body 
decays with $X_u = \pi,\, \eta,\,\eta^\prime,\, \rho,\, \omega$, and decays to nonresonant hadronic final 
states $X_u$.
The branching ratios assumed for the various resonant decays are detailed in Table~\ref{tab:hybridmodel}. 
Exclusive charmless semileptonic decays are simulated using a number of different 
parameterizations: 
for $\Bbar\to \pi\ell\bar{\nu}$ 
decays we use a single-pole ansatz~\cite{BK} for the $q^2$ dependence of the 
form factor with a single parameter measured by \babar~\cite{babarloose};
for decays to pseudoscalar mesons $\eta$ and $\eta'$ and vector mesons 
$\rho$ and $\omega$ we use form factor parameterizations 
based on light-cone sum calculations~\cite{ball04,ball05}.

The simulation of the inclusive charmless semileptonic $\B$ decays to hadronic states with masses 
larger than $2m_{\pi}$  is based on a prescription 
by De~Fazio and Neubert (DFN)~\cite{De Fazio:1999sv} for the
triple-differential decay rate,  $d^3\Gamma\,/\,dq^2\,dE_{\ell}\,ds_H$
($E_{\ell}$ refers to the energy of the charged lepton and $s_H=\mX^2$) with QCD corrections up to ${\cal O}(\alpha_{\rm s})$.
The motion of the $\b$ quark inside the $\B$ meson 
is incorporated in the DFN formalism by convolving the parton-level triple-differential decay rate 
with a non-perturbative shape function (SF). 
This SF describes the distribution of the momentum $k_+$  of the $\b$ quark inside the \B meson.  
The two free parameters of the SF are  ${\bar \Lambda}^\mathrm{SF}$  and  ${\lambda_1}^\mathrm{SF}$. 
The first relates the $B$ meson mass $m_B$ to the $b$ quark mass, 
$m_b^{\mathrm{SF}} = m_B -{\bar \Lambda}^{\mathrm{SF}} $, and ${\lambda_1}^{\mathrm{SF}}$ is the average momentum 
squared of the $b$ quark.
The SF parameterization is of the form $F(k_+) = N (1-x)^a e^{(1+a)x}$, 
where $x = k_+/{\bar \Lambda}^\mathrm{SF} \le 1 $ and $a = -3 ({\bar \Lambda}^\mathrm{SF})^{2}/{\lambda_1}^\mathrm{SF}-1$. 
The first three moments $A_i$ of the SF must satisfy the following relations: $A_0 = 1$, $A_1 = 0$ 
and $A_2 = -{\lambda_1}^\mathrm{SF}/3$.  

The nonresonant hadronic state $X_u$ is simulated with a
continuous invariant mass spectrum according to the DFN prescription.
The fragmentation of the $X_u$ system into final state hadrons 
is performed by {\sc jetset}~\cite{Sjostrand:1994yb}.
The resonant and nonresonant components are combined such that the sum of their branching fractions 
is equal to the measured branching fraction for  inclusive \Bxulnu\ decays~\cite{PDG2010}, 
and the spectra agree with the DFN prediction.
In order to obtain predictions for different values of ${\bar \Lambda}^\mathrm{SF}$ and
${\lambda_1}^\mathrm{SF}$, the generated events are reweighted.
 
\begin{table}[htbp]
\caption{Branching fractions and their uncertainties~\cite{PDG2010} for exclusive $\Bxulnu$ decays }
\begin{center}
\begin{tabular}{lll} \hline\hline
{\bf mode} &\rule{0pt}{9pt}$\BR (\Bzb \to X_u\ell\bar{\nu})$ & $\BR (\Bm\to X_u\ell\bar{\nu})$ \\\hline
\rule{0pt}{9pt}$\Bbar\to \pi\ell\bar{\nu}$           & $(136 \pm 7) \cdot 10^{-6}$  	& $ (77 \pm 12)\cdot 10^{-6}$ \\
$\Bbar\to \eta\ell\bar{\nu}$          &                       		    & $ (64 \pm 20)\cdot 10^{-6}$ \\
$\Bbar\to \rho\ell\bar{\nu}$          & $(247 \pm 33)\cdot 10^{-6}$  	& $(128 \pm 18)\cdot 10^{-6}$ \\
$\Bbar\to \omega\ell\bar{\nu}$        &                      		    & $(115 \pm 17)\cdot 10^{-6}$\\
$\Bbar\to \eta^\prime\ell\bar{\nu}$   &                      		    & $(17 \pm 22)\cdot 10^{-6}$  \\
\hline\hline
\end{tabular}
\end{center}
\label{tab:hybridmodel}
\end{table}
We estimate the shape of background distributions by using
simulations of the process $\epem\to\FourS\to\BB$ with the $B$ mesons
decaying according to measured branching fractions~\cite{PDG2010}. 

For the simulation of the dominant background from \Bxclnu\ decays, we have chosen a variety of 
different form factor parameterizations.
For \Bdlnu and \Bdslnu decays we use 
parameterizations~\cite{Caprini:1997mu} based on heavy quark effective theory 
(HQET)~\cite{Grinstein:1990mj,Eichten:1989zv,Georgi:1990um,Falk:1990yz}.
In the limit of negligible charged lepton masses, decays to pseudoscalar mesons are described by 
a single form factor for which the $q^2$ dependence is expressed in terms of a slope parameter $\rho^2_D$. 
We use the world average $\rho^2_D=1.19 \pm 0.06$~\cite{HFAG2011}, updated with recent precise measurements 
by the \babar\ Collaboration~\cite{Aubert:2008yv,Aubert:2009ac}. 
Decays to vector mesons are described by three form factors, of which the axial 
vector form factor dominates.  In the limit of heavy quark symmetry,
their $q^2$ dependence can be described by three parameters for which we use
the most precise \babar\ measurements~\cite{Aubert:2007rs,Aubert:2008yv}:
$\rho_{D*}^2=1.20 \pm 0.04$~\cite{Aubert:2007rs,Aubert:2008yv}, $R_1=1.429 \pm 0.074$, 
and $R_2=0.827 \pm 0.044$~\cite{Aubert:2007rs}. For the simulation of semileptonic decays to the 
four L=1 charm states, commonly referred to as $D^{**}$ resonances, 
we use calculations of form factors by Leibovich, Ligeti, Stewart, and Wise~\cite{LLSW}. 
We have adopted the prescription by Goity and Roberts~\cite{Goity:1994xn} for nonresonant  
\Bdbsxlnu decays. 

\section{Event Reconstruction and Signal Extraction}
\label{sec:selection}

\subsection{Reconstruction of hadronic \B decays tagging $\BB$ events }
$\Upsilon(4S)\to \BB$ events are tagged by the hadronic decays of one of the $B$ mesons based on a 
semi-exclusive algorithm that was employed in an earlier analysis~\cite{babar_inclusive}.
We look for decays of the type $\breco\rightarrow D^{(*)}Y^{\pm}$, where $D^{(*)}$ is a charmed 
meson ($D^0$, $D^+$, $D^{*0}$, or $D^{*\pm}$) and $Y$ is a charged state decaying to up to five 
charged hadrons, pions or kaons, plus up to two neutral mesons (\KS or \piz). 
The following decay modes of $D$ mesons are reconstructed:
$\Dz \to \Km \pip$, $\Km \pip \piz$, $\Km \pip \pim \pip$, $\KS \pip \pim$ and
$\Dp \to \Km \pip \pip$, $\Km \pip \pip \piz$, $\KS \pip$, $\KS \pip \pip \pim$, $\KS \pip \piz$ with 
$\KS \to \pi^+ \pi^-$.  $D^*$ mesons are identified by their decays, $D^{*+}\to D^0\pi^+$, $D^+\pi^0$, 
and $D^{*0}\to D^0 \pi^0,D^0\gamma$.
Pions and photons from $D^*$ decays are of low energy and therefore the mass difference 
$\Delta M = m(D\pi) - m(D)$ serves as an excellent discriminator for these decays.

Of the 1113 \breco decay chains that we consider, we only retain
the 342 ones with a signal purity  
${\cal P}=S/(S+B)>20\%$, where $S$ and $B$, derived from MC samples, denote the signal and background yields. 
The kinematic consistency of the \breco candidates with $B$ meson decays is checked using \mes
and the energy difference, $\Delta E = (P_B \cdot P_\mathrm{beams} - s/2)/\sqrt{s}$.
We restrict the \breco mass to  $\mes > 5.22$~\gevcc and
require $\Delta E  = 0$~\gev within approximately three standard deviations, where the $\Delta E$ 
resolution  depends on the decay chain.
If an event contains more than one \breco candidate, the decay chain with the highest $\chi^2$ 
probability is chosen. For this purpose we define 
\begin{eqnarray}
\label{for:newchi2}
\chi^2_\mathrm{total} = \chi^2_\mathrm{vertex} + \left( \frac{M_{D^{(*)}_\mathrm{reco}} - M_{D^{(*)}}}{\sigma_{D^{(*)}_\mathrm{reco}}}  \right)^2 + 
\left( \frac{\Delta E }{\sigma_{\Delta E}}  \right)^2. 
\end{eqnarray}
Here the first term 
is taken from a vertex fit for tracks from \breco decays, the second relates 
reconstructed and nominal masses~\cite{PDG2010}, $M_{D^{(*)}_\mathrm{reco}}$ and $M_{D^{(*)}}$, 
of the charm mesons ($D^0,D^+,D^{*0}$ or $D^{*\pm}$), with the resolution 
$\sigma_{D^{(*)}_\mathrm{reco}}$, and the third term checks the 
energy balance $\Delta E$ compared to its resolution $\sigma_{\Delta E}$.
The number of degrees of freedom is therefore defined as  $N^\mathrm{dof} = N^\mathrm{dof}_\mathrm{vertex} +  2$.
The resulting overall tagging efficiency is $0.3\%$ for $\BzBzb$ and $0.5\%$ for \BpBm events.

\subsection{Selection of inclusive \Bxlnu decays}
In order to minimize systematic uncertainties,
we measure the yield of selected charmless semileptonic decays in a specific kinematic region  
normalized to the total yield of semileptonic \Bxlnu decays. Both semileptonic decays, 
the charmless and the normalization modes, are identified by at least one charged lepton
in events that are tagged by a \breco decay. 
Both samples are background-subtracted and corrected for efficiency. 
Using this normalization, the systematic uncertainties on the \breco reconstruction and 
the charged lepton detection cancel in the ratio or are eliminated to a large degree.

The selection criteria for the charmless and the total semileptonic samples
are chosen to minimize the statistical uncertainty of the 
measurement as estimated from a sample of fully simulated MC events that includes both signal and 
background processes.

A restriction on the momentum of the electron or muon is applied 
to suppress backgrounds from secondary charm or 
$\tau^{\pm}$ decays, photon conversions and misidentified hadrons. 
This is applied to \Pl, the lepton momentum in the rest frame of the recoiling \B meson, 
which is accessible since the momenta of the \FourS and the reconstructed \B are known.  
This transformation is important because theoretical calculations refer to variables that are 
Lorentz-invariant or measured in the rest frame of the decaying $B$ meson. 
We require \Pl to be greater than 1~\gevc, for which about 90\% of the signal is retained.

For electrons and muons the angular acceptance is defined as $0.450 < \theta < 2.473$~rad, where $\theta$ 
refers to the polar angle relative to the electron beam in the laboratory frame.
This requirement excludes regions where charged particle tracking and identification 
are not efficient. We suppress muons from $J/\psi$ decays
by rejecting the event if a muon candidate paired with any other charged track of opposite charge 
(and not part of \breco) results in an invariant mass of the pair that is consistent with the $J/\psi$ mass.
A similar requirement is not imposed on electron candidates, because of the 
poor resolution of the corresponding $J/\psi$ peak.

We also reject events if the electron candidate paired with any other charged track 
of opposite charge is consistent with a $\gamma \to e^+e^-$ conversion.

A variety of processes contributes to the inclusive semileptonic event samples, 
${\it i.e.}$ candidates selected by a \breco
decay and the presence of a high momentum lepton. In addition to true semileptonic
decays tagged by a correctly reconstructed \breco, we consider the following classes of backgrounds:
\begin{itemize}
\item {\it Combinatorial background:} 
the \breco\ is not correctly reconstructed. 
This background originates from \BB\ or  continuum $e^+e^-\to q{\bar q}(\gamma)$ events.
In order to subtract this background, the yield of true \breco\ decays is determined from 
an unbinned maximum likelihood fit to the \mes distribution (Section~\ref{sec:mesfit}). 
\item {\it Cascade background:}
  the lepton does not originate from a semileptonic \B decay, but 
  from secondary  decays, for instance from $D$ mesons, including $D_s \to \tau \nu$, 
  or residual $J/\psi$ background.
\item {\it $\tau$  background:} electrons or muons originate from prompt $\tau$ leptons, 
  primarily from $\Bbar \to X \tau \bar{\nu}$ decays.
\item {\it Fake leptons:} hadrons are misidentified as leptons, primarily muons. 
\end{itemize}
The last three sources of background are combined and in the following are referred to as ``other'' background.

\subsection{Selection of inclusive \Bxulnu decays}
\label{sigselection}
A large fraction of \Bxclnu decays is expected to have a second lepton from cascade decays of the 
charm particles. In contrast, in \Bxulnu decays secondary leptons are very rare. 
Therefore, we enhance signal events by selecting events with only one charged lepton having $\Pl>1$~\gevc.

In semileptonic \B meson decays, the charge of the primary lepton is equal to the sign of the 
charge of the \b quark.  Thus for \BpBm\ events in which the \breco and the lepton originate from 
different \B decays in the event, we impose the requirement $Q_{b} Q_\ell<0$, where $Q_{b}$ is the 
charge of the \b quark of the \breco\ and $Q_\ell$ is the charge of the lepton. 
For $\BzBzb$ events this condition does not strictly hold because of flavor mixing. 
Thus, to avoid a loss in efficiency, this requirement is not imposed.
The hadronic state $X_u$ in charmless semileptonic decays is reconstructed from all particles 
that are not associated with the \breco\ candidate or the charged lepton.
The measured four-momentum $P_X$ is defined as
\begin{equation}
P_X=\sum_{i=1}^{N_\mathrm{trk}}P_i^\mathrm{trk} + \sum_{i=1}^{N_{\gamma}}P_i^{\gamma},\\
\end{equation}
where the summation extends over the four-vectors of the charged particles and photon candidates.
From this four-vector, other kinematic variables, $\mx^2 =P_X^2=E_X^2 - p_X^2$, 
$\Q = P_{B_{\mathrm{reco}}}-P_X$ ($P_{B_{\mathrm{reco}}}$ being the \breco four-momentum), and $P_+$, can be calculated.
The loss of one or more charged or neutral particles or the addition of tracks or single 
electrons from photon conversions degrade the reconstruction of $X_u$ and the resolution of the 
measurement of any related kinematic variables.
In order to reduce the impact of missing charged particles and the effect of single electrons from 
$\gamma\to e^+ e^-$ conversions, we impose charge conservation on the whole event,
$Q_\mathrm{tot}=Q_{\breco}+Q_{X}+Q_{\ell}=0$.
This requirement rejects a larger fraction of \Bxclnu events because of 
their higher charged multiplicity and the presence of very low momentum charged pions 
from $D^{*\pm}\to \D^0 \pi^{\pm}_\mathrm{soft}$ decays which have low detection efficiency. 

In \Bxlnu decays, where the state $X$ decays hadronically, the only undetected particle is a neutrino.
The neutrino four-momentum $P_{\nu}$ can be estimated from the missing momentum four-vector 
$P_\mathrm{miss} = P_{\FourS}-P_{\breco}-P_X-P_\ell$.
For correctly reconstructed events with a single semileptonic decay, the missing mass squared, 
$\mathit{MM}^2=P_\mathrm{miss}^2$, is consistent with zero. Failure to detect one or more 
particles in the event creates a tail at large positive values; thus $\mathit{MM}^2$ is used as a 
measure of the quality of the event reconstruction. 
Though $\mathit{MM}^2$ is Lorentz invariant, the missing momentum is usually measured in the laboratory 
frame, because this avoids the additional uncertainty related to the 
transformation into the c.m. frame. We require $\mathit{MM}^2$ to be less than 0.5~\gevccsq.  
Because of the higher probability for additional unreconstructed neutral particles, a neutrino 
or $K_L$, the $\mathit{MM}^2$ distribution is broader for \Bxclnu decays, and this restriction 
suppresses this background more than signal events.
 
In addition, we suppress the \Bdslnu background by exploiting the small $Q$-value of the 
$D^*\to D\pi_{\mathrm{soft}}$ decays which result in a very low momentum pion. 
For energetic $D^*$ mesons, the momenta  
$p_{\pi_\mathrm{soft}}$ and $p_{D}$  are almost collinear, and 
we can approximate the $D^*$ direction by the $\pi_\mathrm{soft}$ direction and estimate  
the $D^*$ energy by a simple approximation based on the $E_{\pi_\mathrm{soft}}$,  
$E_{D^*}\approx m_{D^*}\times E_{\pi_\mathrm{soft}}/145$~\mevcc. 
Using the measured \breco\ and charged lepton momenta, and the four-momentum of the 
$D^*$ derived from any pion with c.m. momentum below $200$~\mevc, we estimate the 
neutrino mass for a potential 
\Bdslnu decay as $\mathit{MM}^2_\mathrm{veto}=(P_B-P_{D^*}-P_\ell)^2$.
For true \Bdslnu decays, this distribution peaks at zero.  
Thus, we veto $D^*$ decays to low momentum charged or neutral pions by requiring, respectively,
$\mathit{MM}^2_\mathrm{veto}(\pi^+_\mathrm{soft})<-3$~\gevccsq or 
$\mathit{MM}^2_\mathrm{veto}(\pi^0_\mathrm{soft})<-2$~\gevccsq.
This is achieved without explicit reconstruction of the $D$ meson decays, and thus avoids large 
losses in rejection power for this veto.

We reduce \Bdslnu background by vetoing events with a charged or neutral 
kaon ($\KS\to\pi^+\pi^{-}$), that originate primarily from the decays of  charm particles.  

\begin{figure*}[htb]
 \begin{centering}
 \includegraphics[width=0.45\textwidth]{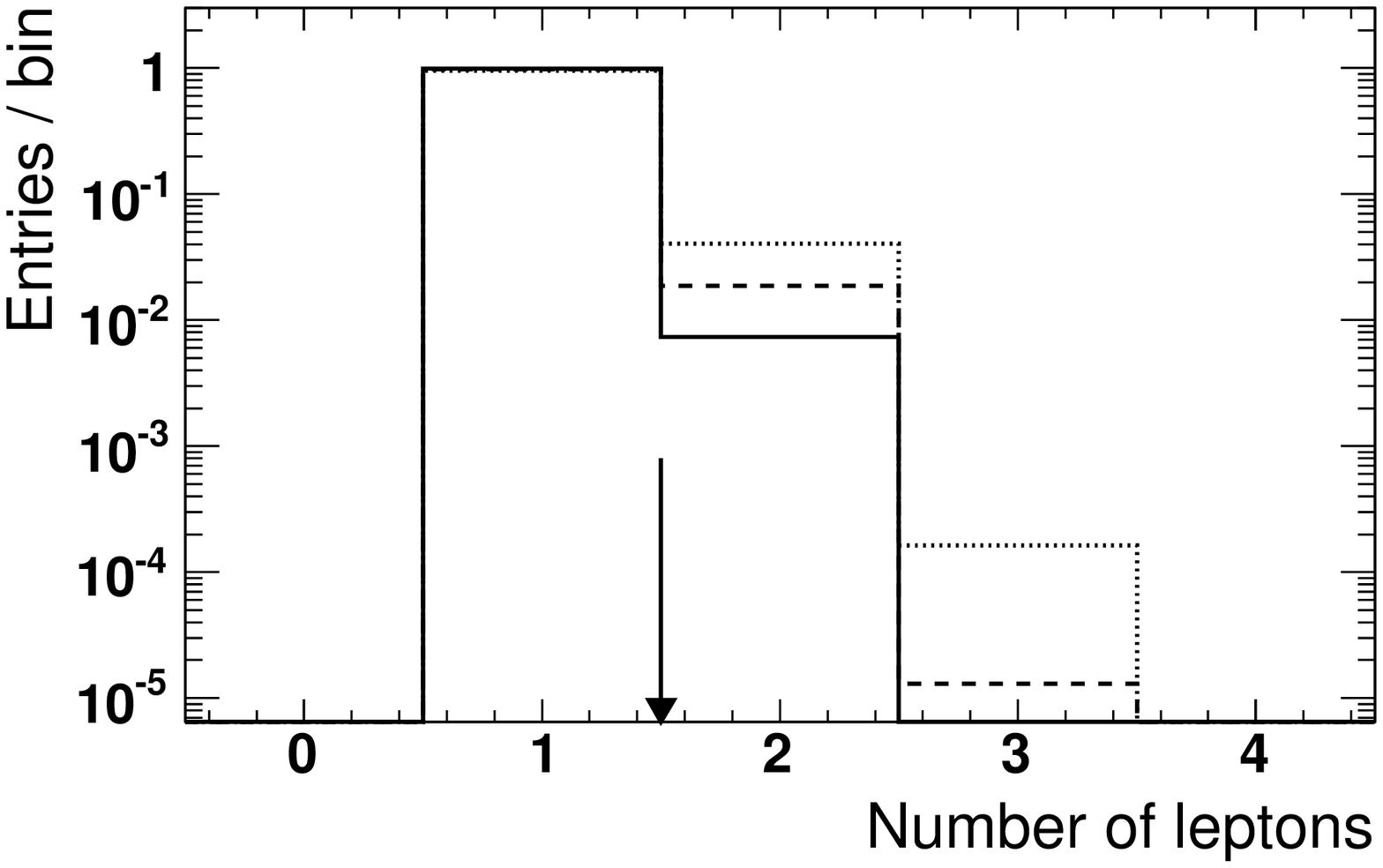} 
 \includegraphics[width=0.45\textwidth]{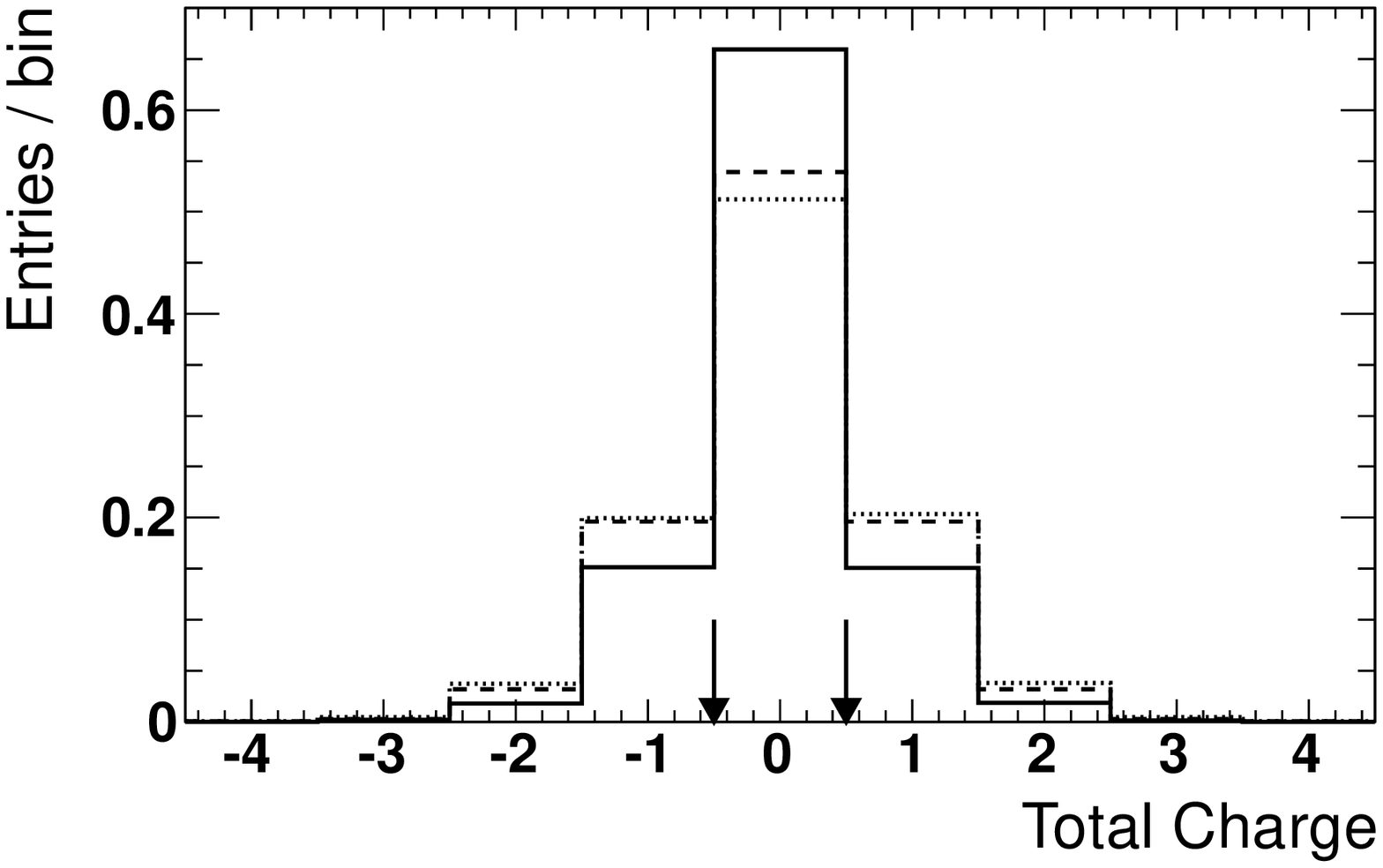} 
 \includegraphics[width=0.45\textwidth]{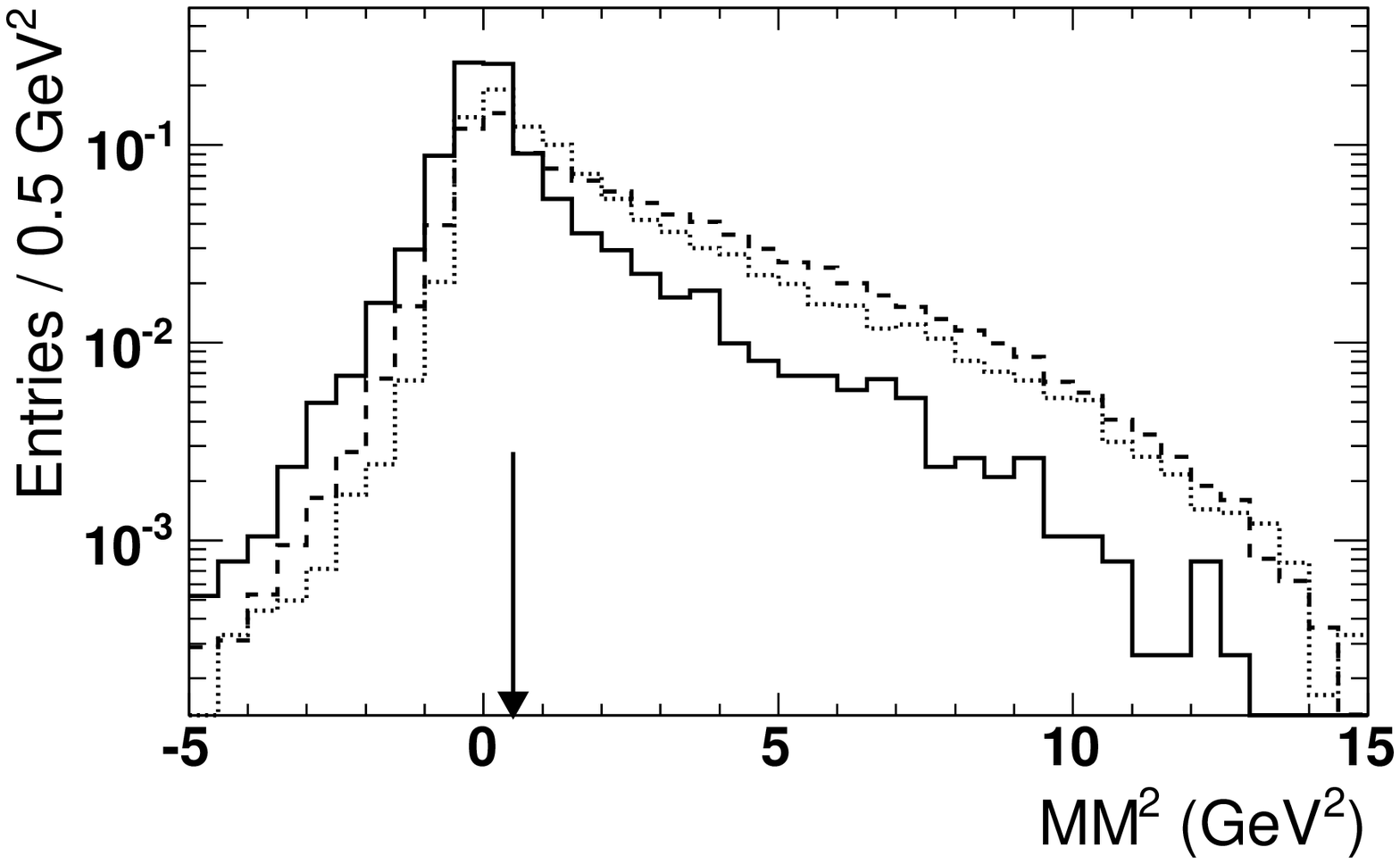} 
 \includegraphics[width=0.45\textwidth]{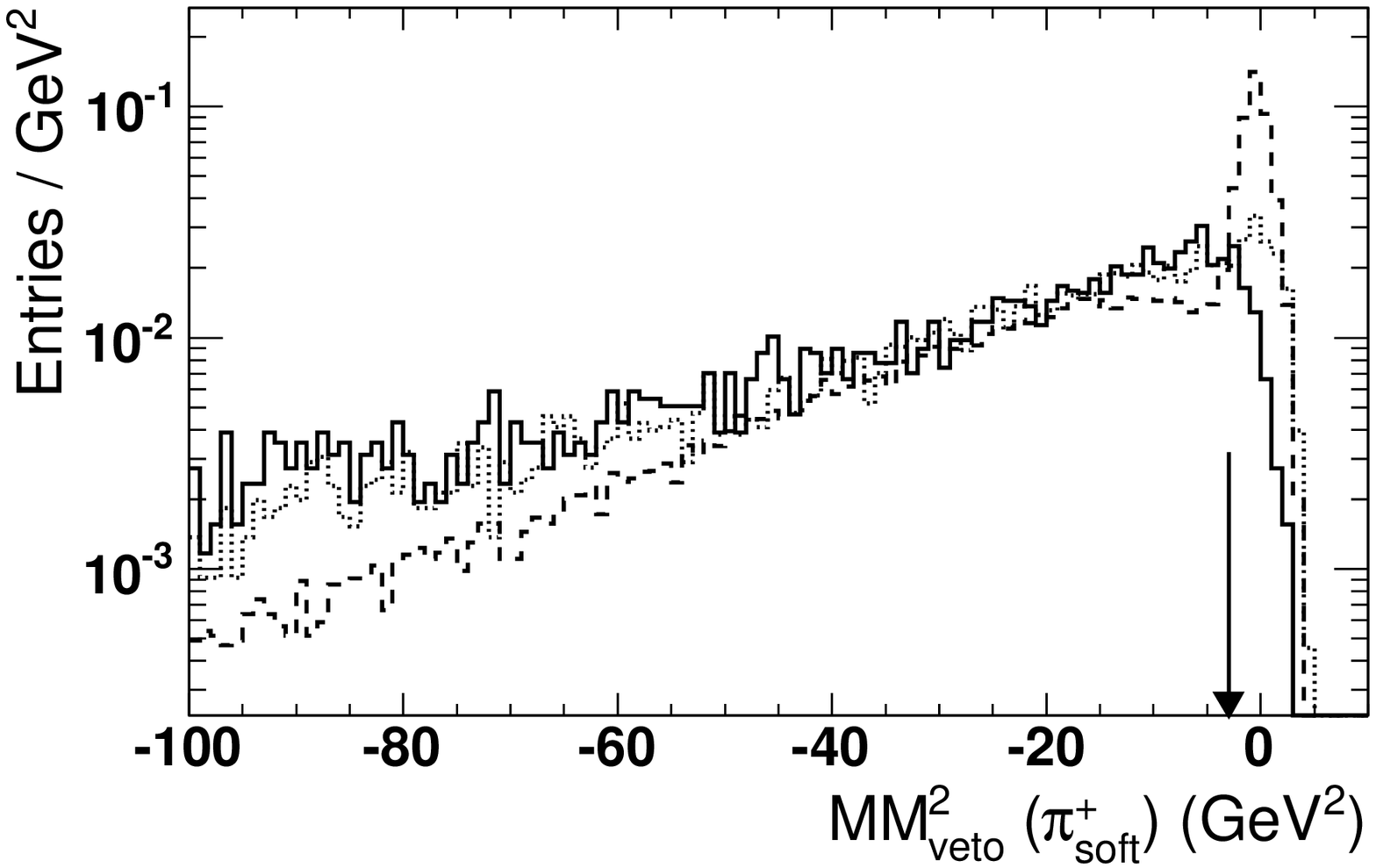} 
 \includegraphics[width=0.45\textwidth]{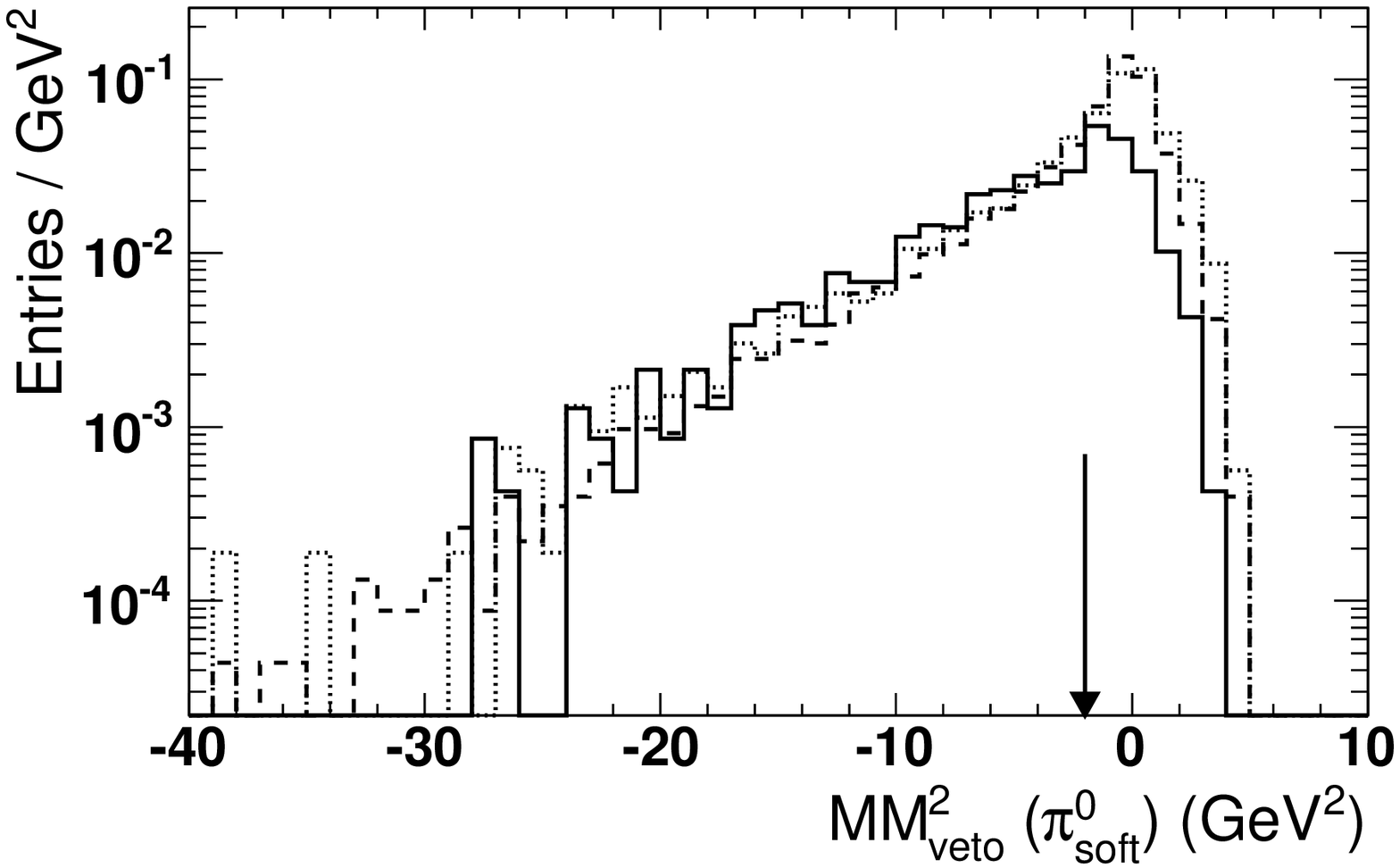} 
 \includegraphics[width=0.45\textwidth]{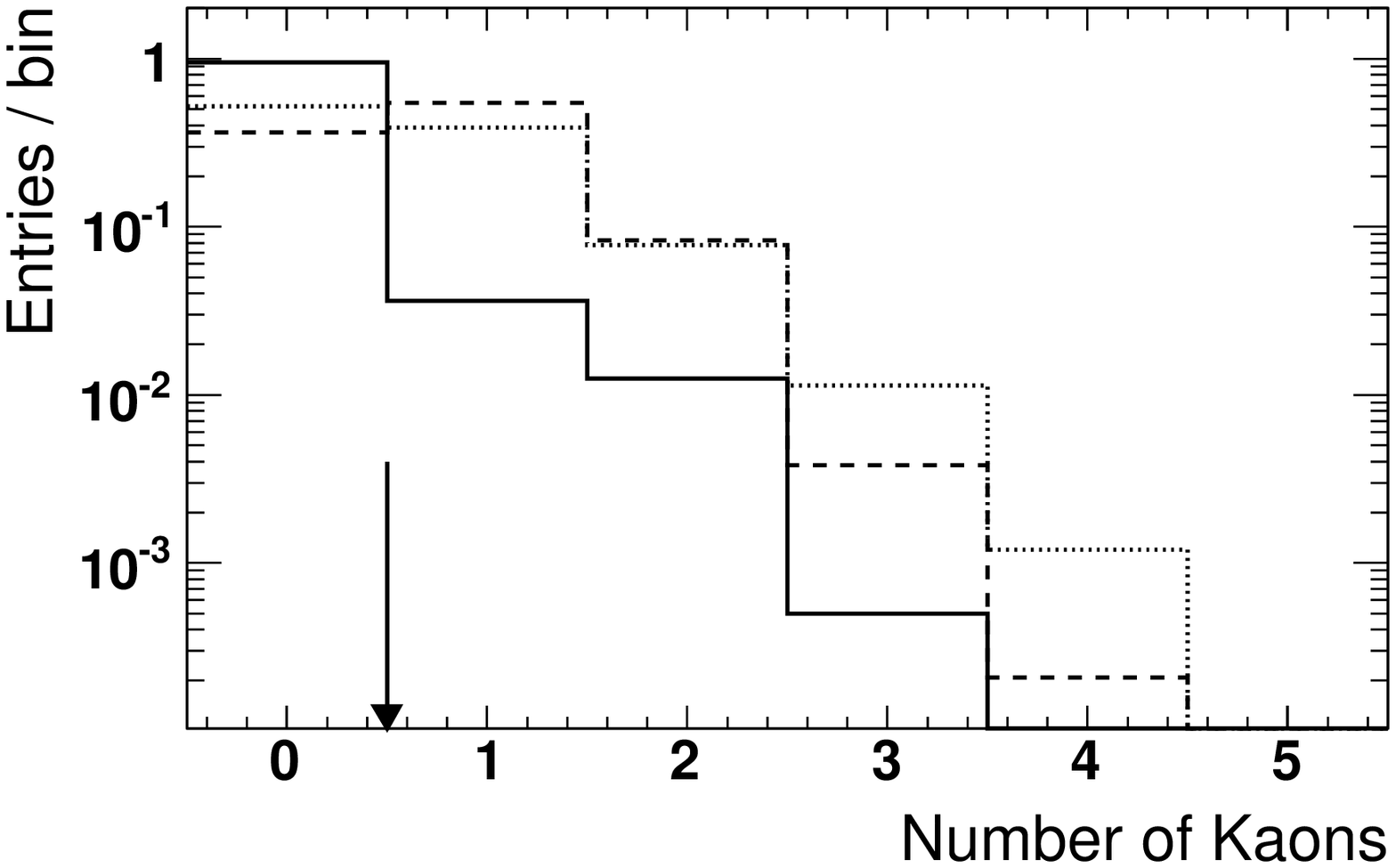} 
\caption{MC distribution of the kinematic variables for which we 
  apply restrictions sequentially as listed in Table~\ref{tab:seleff}, 
  for \Bxulnu (solid line), \Bxclnu (dashed
  line) and ``other'' component (dotted line). All distributions are
  normalized to unity and selection criteria have been applied
  cumulatively, except those affecting directly the variable shown. 
  The arrows indicate the selection
  requirement for a specific variable, as described in
  Section~\ref{sigselection}.  
\label{fig:kinvar}}
 \end{centering}
\end{figure*}

A summary of the impact of the signal selection criteria on the high-energy lepton sample, for
the signal, semileptonic and nonsemileptonic background samples is presented in Table~\ref{tab:seleff},
in terms of cumulative selection efficiencies. 
Figure~\ref{fig:kinvar} shows the kinematic variables that
appear in Table~\ref{tab:seleff} for different event categories.
Combinatorial background is not included; it is subtracted 
based on fits to the \mes\ distributions, as described in Section~\ref{sec:mesfit}. 
The overall efficiency for selecting charmless semileptonic decays in the sample of tagged events with a 
charged lepton is $33.8\%$; the background reduction is 97.8\% for \Bxclnu and 95.3\% for ``other''.

\begin{figure*}[htb!]
    \begin{centering}
    \begin{tabular}{c c c}
      \includegraphics[width=0.325\textwidth]{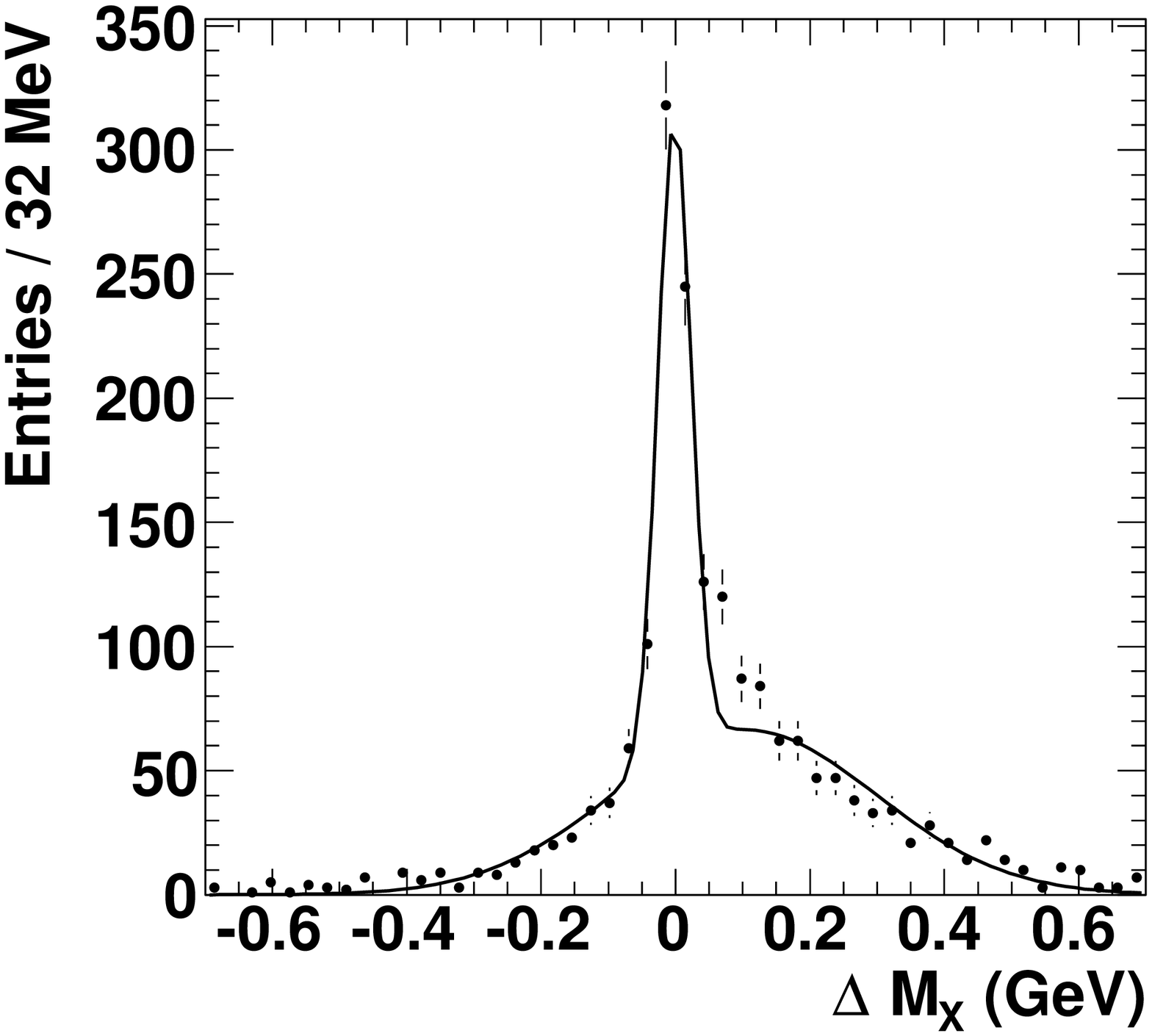}&
      \includegraphics[width=0.325\textwidth]{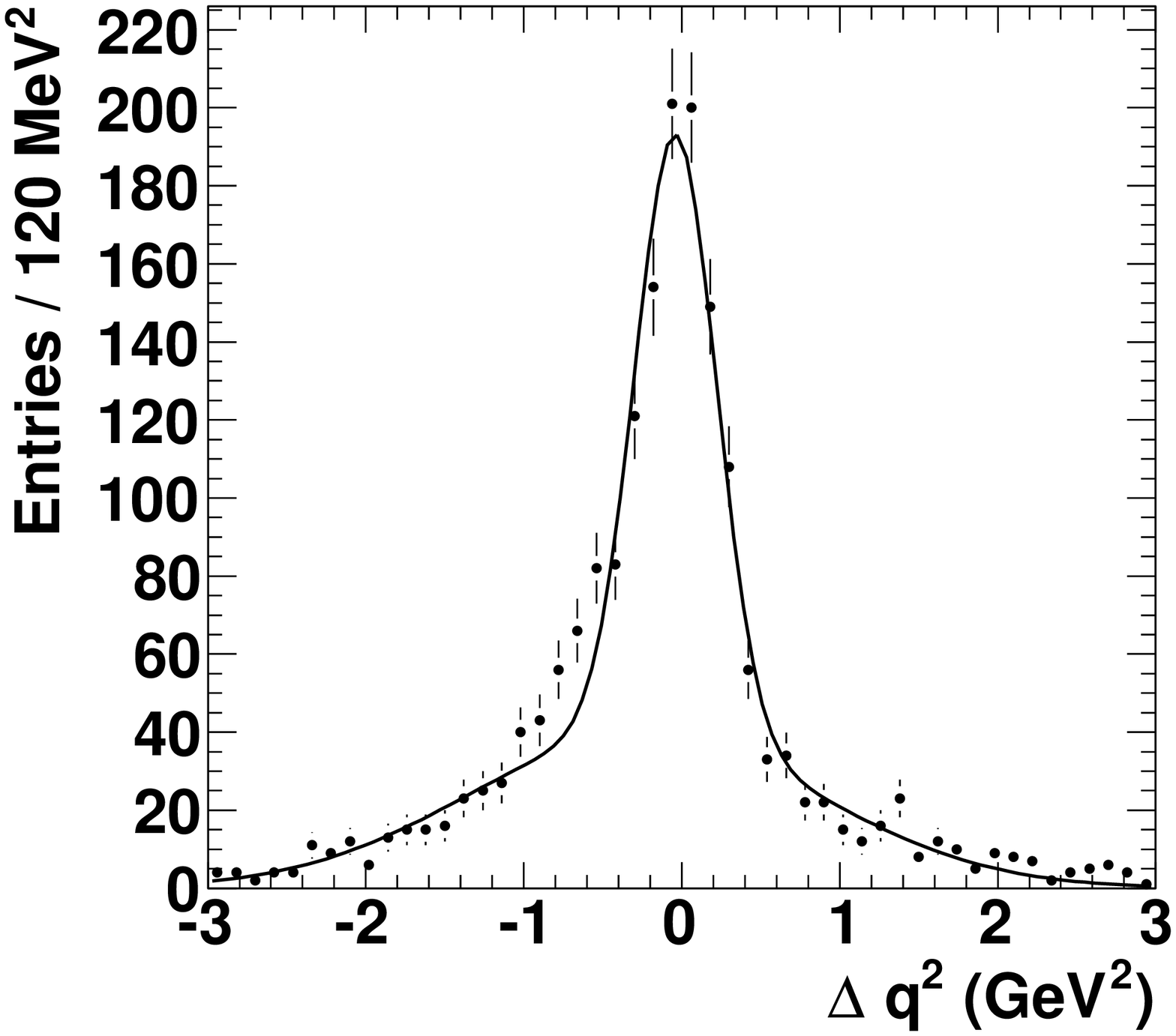}&
      \includegraphics[width=0.325\textwidth]{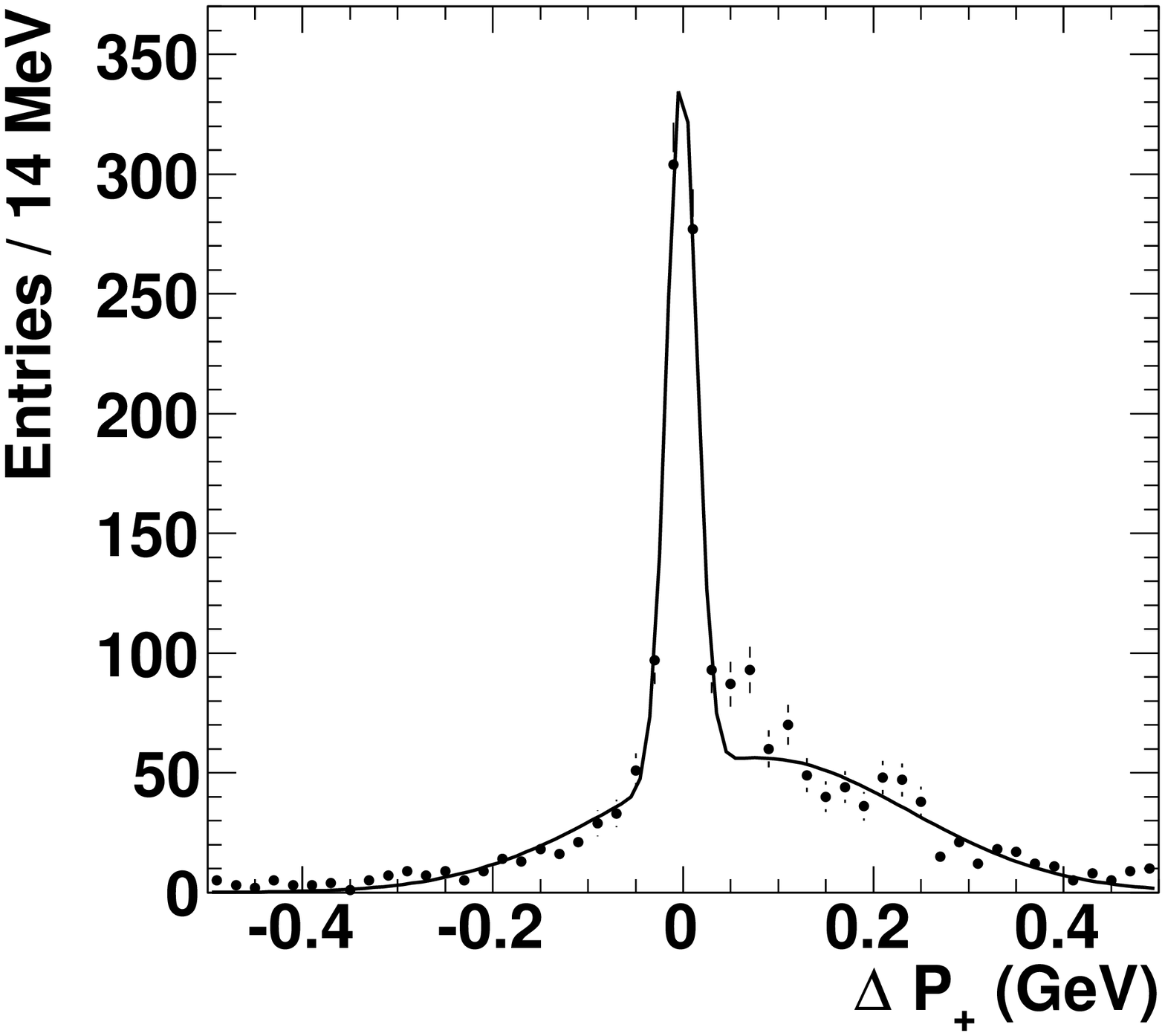} \\
      \end{tabular}
	  \caption{Resolution for MC simulated for signal \Bxulnu events passing all event selection criteria, 
	(left) ${\mX}_\mathrm{reco}-{\mX}_\mathrm{true}$,  
	(center) ${\Q}_\mathrm{reco}-{\Q}_\mathrm{true}$, and (right) $P_{+,\mathrm{reco}}-P_{+,\mathrm{true}}$. 
	The curve shows a fit  results for the sum of two Gaussian functions.}
	  \label{fig:resolution}
   \end{centering}
   \end{figure*}

The resolution functions determined from MC simulation of signal events passing the selection 
requirements are shown in Fig.~\ref{fig:resolution} for the variables \mx, \Q, and $P_+$.
Each of these distributions has a narrow core containing
30\%, 50\%, and 30\% of the \Bxulnu events, 
with widths of ~$25\mevcc$, $250\mev^2$, and $10 \mevc$, 
respectively. The remaining events have a considerably poorer resolution,  primarily because of 
lost secondary particles from the decay of the hadronic $X_u$.

\begin{table}[htb]
\caption {\label{tab:seleff}  
Comparison of the cumulative selection efficiencies for samples of signal \Bxulnu
decays and \Bxclnu and ``other'' backgrounds. 
The efficiencies are relative to the sample of \breco -tagged events with a charged lepton.}
\begin{tabular}{l  c c c }
\hline\hline
Selection  & \rule{0pt}{9pt} \Bxulnu & \Bxclnu & Other \\ \hline
Only one lepton             & 99.3\% & 98.1\% & 95.8\% \\
Total charge Q=0			& 65.5\% & 52.9\% & 49.1\% \\
$\mathit{MM}^2$ 			& 44.2\% & 17.8\% & 17.8\% \\
$D^*\ell\bar{\nu}~(\pi^+_s)$ veto 	& 40.6\% & 9.9\% & 14.4\% \\
$D^*\ell\bar{\nu}~(\pi^0_s)$ veto 	& 34.8\% &  6.3\% &  9.1\% \\
Kaon veto 				& 33.8\% &  2.2\% &  4.7\% \\
\hline\hline
\end{tabular}
\end{table}

On the basis of the kaon and the $D^*$ veto, two data samples are defined:
\begin{itemize}
\item {\it{signal-enriched:}} events that pass the vetoes; this sample is used to extract the signal;
\item {\it{signal-depleted:}} events rejected by at least one veto; they are used 
as control sample to check the agreement between data and simulated  backgrounds,
including the poorly understood \Bdsslnu decays.
\end{itemize}

\subsection{Subtraction of combinatorial background}
\label{sec:mesfit}
The subtraction of the combinatorial background of the \breco tag
for the signal and normalization samples relies on unbinned maximum-likelihood fits to the \mes
distributions. 
For signal decays the goal is to extract the distributions in the kinematic 
variables \Pl , \mx, \Q, and $P_+$. Because the shapes and relative yields of the signal and 
background contributions depend on the values of these kinematic variables, 
the continuum and combinatorial background subtraction is performed separately for 
subsamples corresponding to events in  bins of these variables.
This results in  more accurate spectra than a single fit to the full sample of events in each 
selected region of phase space. 

For the normalization sample, the fit is performed for the full event sample,
separately for \Bzb and \Bm tags.

The \mes distribution for the combinatorial \breco background can be described 
by an ARGUS function~\cite{argusf},
\begin{equation}
  \label{eq:argus}
  f_\mathrm{bkg}(m) = N_\mathrm{bkg} m \sqrt{1-m^2} e^{-\xi(1-m^2)},
\end{equation}
where $m=\mes/\mes^\mathrm{max}$ and $\mes^\mathrm{max}$ is the endpoint of the \mes distribution
which depends on the beam energy, and $\xi$ determines the shape of the function. 
$N_\mathrm{bkg}$ refers to the total number of background events in the distribution.

For signal events, the \mes distribution resembles a resolution function peaking at the $B$ meson 
mass with a slight tail to lower masses. Usually the peak of the \mes\ distribution is empirically described by 
a Crystal Ball function~\cite{cristal-ball}, but this ansatz turned out to be inadequate for this 
dataset because the \breco sample is composed of many individual decay modes with different 
resolutions. We therefore follow an approach previously used in \babar\ data~\cite{Aubert:2006au} 
and build a more general function, using a Gaussian function, $f_g(x) = e^{-{x^2}/2}$, and the 
derivative of $\tanh{x}$, $f_t(x) = e^{-x}/(1+e^{-x})$, to arrive at
\begin{equation}
  \label{eq:signal}
  f_{sig}(\Delta) =\begin{cases}
  			\frac{C_2}{(C_3-\Delta)^n}  & \mathrm{if}\ \Delta < \alpha \\
            \frac{C_1}{\sigma_L}f_{t}(\frac{\Delta}{\sigma_L})   & \mathrm{if}\ \alpha \leqslant \Delta < 0 \\ 
            \frac{r}{\sigma_1}  f_{t}(\frac{\Delta}{\sigma_1}) + \frac{1-r}{\sigma_2} f_g (\frac{\Delta}{\sigma_2}) & \mathrm{if}\ \Delta \geqslant 0.
              \end{cases}
\end{equation}
Here $\Delta = \mes - \overline{m}_{\rm ES}$, where $\overline{m}_{\rm ES}$ is the maximum of 
the \mes distribution. $C_1, C_2$ and $C_3$ are functions of the parameters $\overline{m}_{\rm ES}$, $r, 
\sigma_1, \sigma_2, \sigma_L, \alpha$, and $n$, that ensure the continuity of $f_{sig}$.

\begin{figure}[htb]
  \begin{center}
    \includegraphics[width=0.48\textwidth]{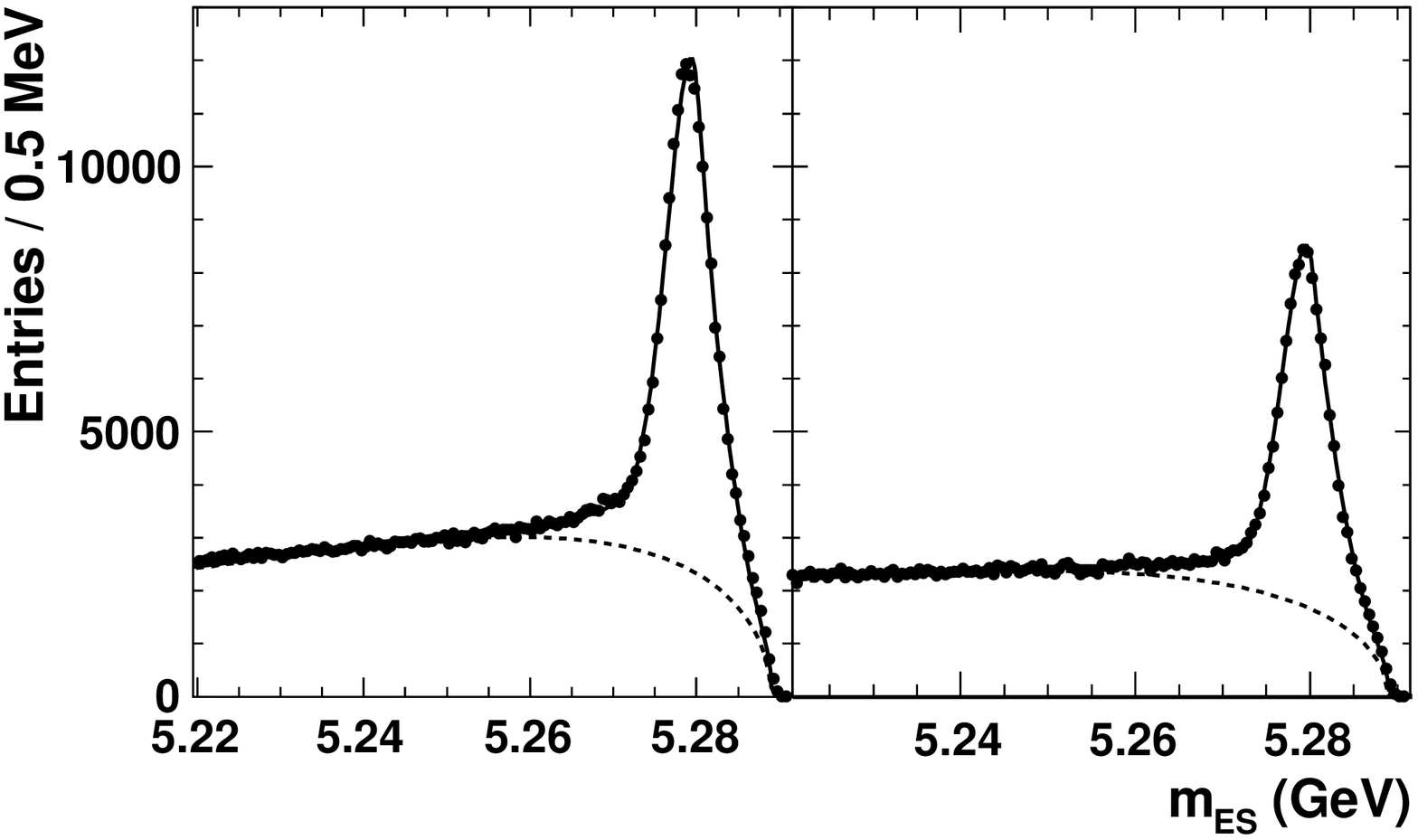}
    \caption{The \mes distribution for the inclusive
	 semileptonic sample, for fully reconstructed hadronic decays of \Bm (left) and \Bzb mesons (right). 
	 The solid line shows the result of the 
     maximum-likelihood fit to signal and combinatorial backgrounds; 
     the dashed line indicates the shape of the background described by an ARGUS function.}
    \label{fig:mes:data}
  \end{center}
\end{figure}

Given the very large number of parameters, we first perform a fit to samples
covering the full kinematic range and determine all parameters describing  
$f_{sig}$ and the ARGUS function.  We then repeat the fit for events in each bin of the 
kinematic variables, with only the relative normalization of the signal and 
background, and the shape parameter $\xi$ of the ARGUS function as free parameters. 
Figure~\ref{fig:mes:data} shows the \mes distribution for the inclusive semileptonic sample, separately 
for charged and neutral \B mesons.

Finally, we correct for the contamination from cascade background in the number of 
neutral \B mesons, due to the effect of \Bz-\Bzb mixing, in each bin of the kinematic variables. 
We distinguish neutral \B decays with right- and wrong-sign leptons, 
based on the flavor of the \breco decay.
The contribution from cascade decays is subtracted by computing the number of neutral 
\B mesons $N_{\Bz}$ as
\begin{equation}
N_{\Bz} = \frac{1-\chi_d}{1-2\chi_d}N_{B^0_\mathrm{rs}}-\frac{\chi_d}{1-2\chi_d}N_{B^0_\mathrm{ws}},
\end{equation}
where $N_{B^{0}_\mathrm{rs}}$ and $N_{B^0_\mathrm{ws}}$ are the number of neutral 
\B mesons with right and wrong sign of the charge of the accompanying lepton, and
$\chi_d=0.188\pm 0.002$~\cite{PDG2010} is the \Bz-\Bzb mixing parameter.

The performance of the \mes fit has been verified using MC simulated distributions. 
We split the full sample in two parts.
One part, containing one third of the events, is treated as data, and is similar in size  to the
total data sample.
The remaining two thirds represent the simulation. 
The fit procedure, described in Section~\ref{meastec}, is applied to these samples and
yields, within uncertainties, the charmless semileptonic branching fraction that is input to the
MC generation.

\section{Signal extraction and partial branching fraction measurement}
\label{meastec}
\subsection{Signal yield}
\begin{figure*}[htb]
    \begin{centering}
      \includegraphics[width=\textwidth]{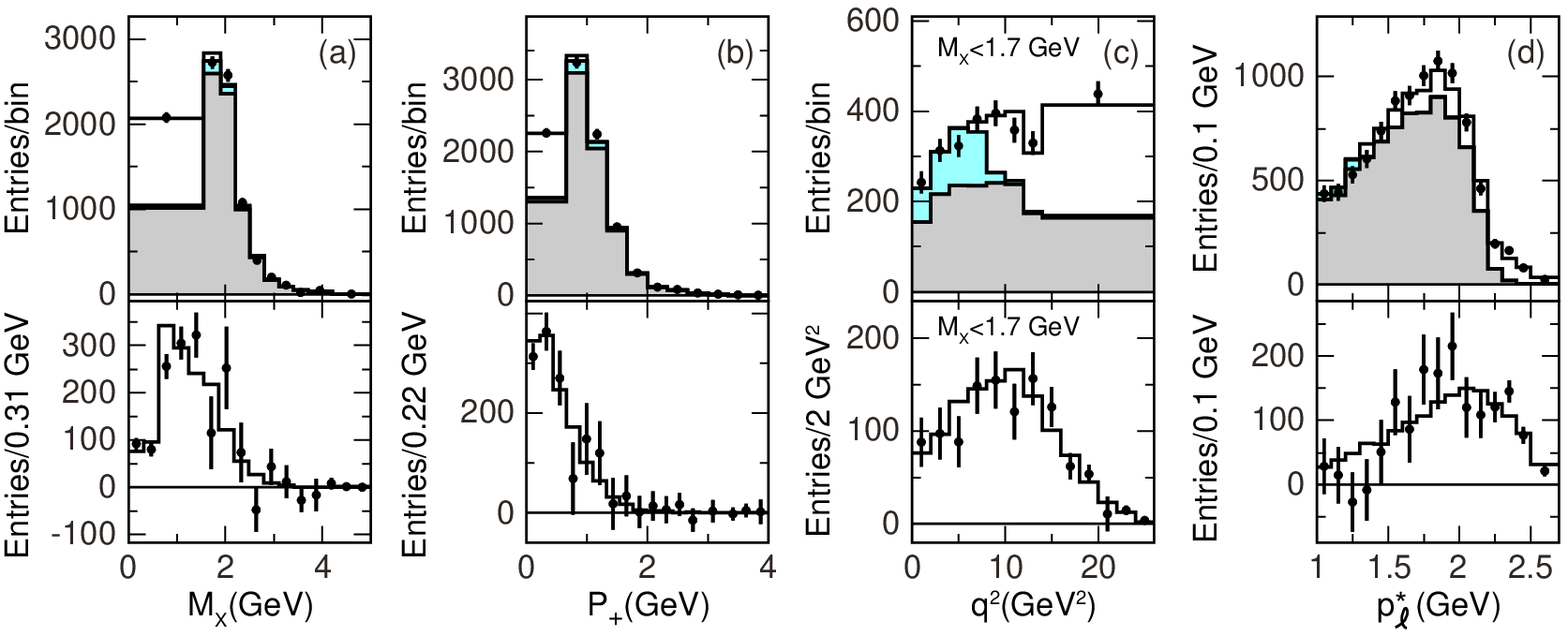}
      \caption{
      Measured distributions (data points) 
      of (a) \mX,  (b) \Pplus, (c) \Q\ with $\mX<1.7$~\gevcc, and (d) $\Pl$.
      Upper row: comparison with the result of the $\chi^2$ 
      fit with varying bin size for the sum of two scaled MC contributions 
      (histograms), \Bxulnu decays generated inside (white) or outside 
      (light shading) the selected kinematic region, 
      and the background  (dark shading).  
      Lower row: corresponding spectra with equal bin size after 
      background subtraction based on the fit.
      The data are not corrected for efficiency.}
    \label{fig:results}
   \end{centering}
   \end{figure*}

\begin{figure*}[htb]
\begin{centering}
 \includegraphics[width=0.7\textwidth]{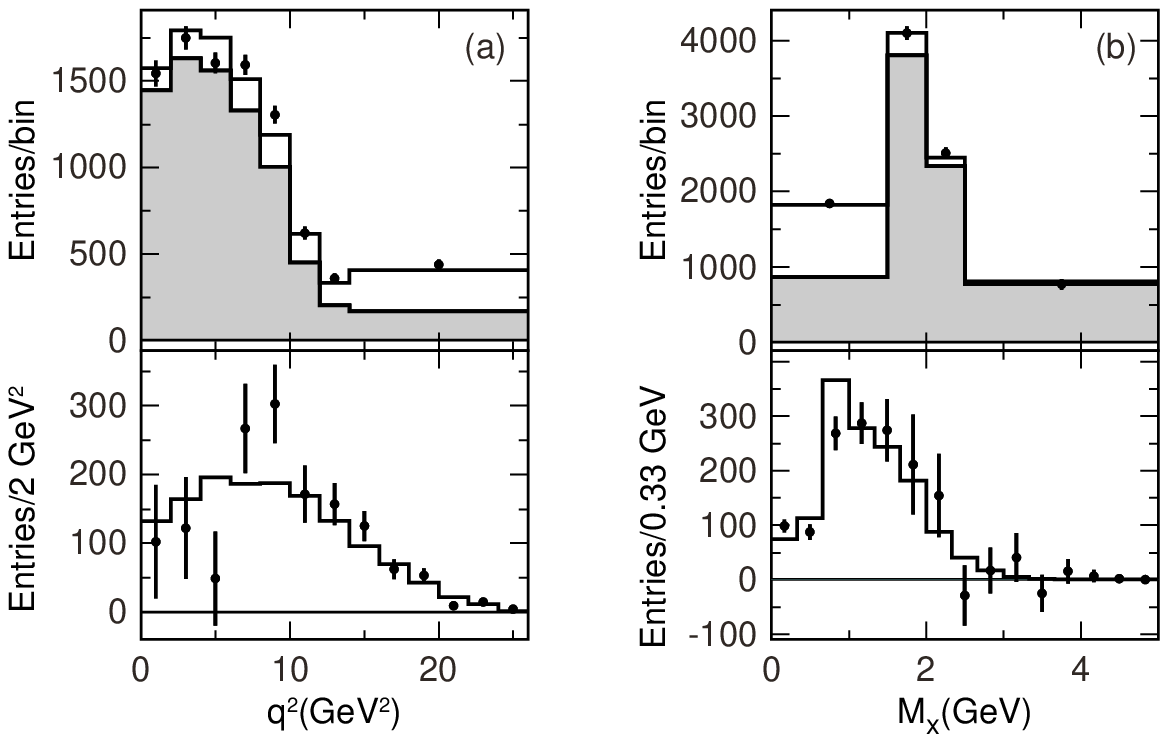} 
\caption{
Projections of measured distributions (data points) 
      of (a) \Q and (b) \mX with varying bin size,
      for the fit to the \mX\ -- \Q distribution without
      constraints other than $\Pl>1$~\gev.
      Upper row: comparison with the result of the $\chi^2$ 
      fit to the two-dimensional \mX\ -- \Q distribution
      for the sum of two scaled MC contributions (histograms), 
      \Bxulnu\ decays (white) and the background  (dark shading).  
      Lower row: corresponding spectra with equal bin size after 
      background subtraction based on the fit.
      The data are not corrected for efficiency.
}
\label{fig:results_full}
\end{centering}
\end{figure*}

Once continuum and combinatorial \BB\ backgrounds have been
subtracted and the mixing correction has been applied,
the resulting differential distributions of the kinematic variables
are fitted using a $\chi^2$ minimization to extract $N_u$, 
the number of selected signal events.
The $\chi^2$ for these fits is defined as
\begin{equation}
\label{eq:chi2fit}
  \chi^2=
  \sum_i  \frac{[N^{i} - (C_{sig}N_u^{i,\mathrm{MC}} + C_\mathrm{bkg}N_\mathrm{bkg}^{i,\mathrm{MC}})]^2}
{\sigma(N^{i})^2+\sigma(N^{i,\mathrm{MC}})^2},\\
\end{equation}
where, for each bin $i$ of variable width, $N^{i}$ is the number of observed 
events, and $N_u^{i,\mathrm{MC}}$ and $N_\mathrm{bkg}^{i,\mathrm{MC}}$ are the 
number of MC predicted events for signal and background, respectively.
The statistical uncertainties
$\sigma(N^{i})$ and $\sigma(N^{i,\mathrm{MC}})$ are 
are taken from fits to the \mes\ distributions in data and MC simulations.
The scale factors  
$C_\mathrm{sig}$ and $C_\mathrm{bkg}$ are free parameters of the fit. 
The differential distributions are compared with the sum of the signal and background 
distributions resulting from the fit in Figs.~\ref{fig:results} and~\ref{fig:results_full}. 
For the \Bxulnu signal contributions we distinguish between decays that were generated 
with values of the kinematic variable inside the restricted phase space regions,
and a small number of events, $N_u^\mathrm{out}$,
with values outside these regions. This distinction allows us to relate the 
fitted signal yields to the theoretical calculations applied to extract \Vub. 

\subsection{Partial branching fractions}
We obtain partial branching fractions for charmless semileptonic decays 
from the observed number of signal events in the kinematic regions considered, 
after correction for background and efficiency, and
normalization to the total number of semileptonic decays \Bxlnu observed in the 
\breco event sample. 
For each of the restricted regions of phase space under study, we calculate the ratio 
\begin{eqnarray}
 \Delta R_\mathrm{u/sl} 
&=&
 \frac{\Delta\BR({\Bxulnu})}{\BR(\Bxlnu)}=
 \frac{N_u^\mathrm{true}}{N_\mathrm{sl}^\mathrm{true}} \nonumber \\ 
&=& \frac{(N_u)/(\epsilon_\mathrm{sel}^u \epsilon_\mathrm{kin}^u)}{(N_\mathrm{sl}-BG_\mathrm{sl})} 
 \; \frac{\epsilon_\ell^\mathrm{sl} \epsilon_\mathrm{tag}^\mathrm{sl} }{\epsilon_\ell^u \epsilon_\mathrm{tag}^u }.
\label{eq:ratioBR}
\end{eqnarray}
Here, $N_{u}^\mathrm{true}$ and $N_\mathrm{sl}^\mathrm{true}$ refer to the true number of signal
and normalization events. 
The observed signal yield $N_u$ is related to $N_{u}^\mathrm{true}$ 
by $N_u=\epsilon_\mathrm{sel}^u \epsilon_\mathrm{kin}^u \epsilon_l^u \epsilon^u_\mathrm{tag} N_u^\mathrm{true}$,
where $\epsilon_\mathrm{sel}^u$ is the efficiency for detecting \Bxulnu decays in the tagged 
sample after applying all selection criteria, $\epsilon_\mathrm{kin}$ 
is the fraction of signal events with both true and reconstructed
\mX, $P_+$, \Q, or $\Pl$ within the restricted region of phase space, 
and $\epsilon_l^{u}$ refers to the efficiency for selecting 
a lepton from a \Bxulnu decay with a momentum $\Pl>1$~\gevc in a signal event tagged 
with efficiency $\epsilon^{u}_\mathrm{tag}$.
Similarly, $N_\mathrm{sl}^\mathrm{true}$ is related to 
$N_\mathrm{sl}$, the fitted number of observed \breco accompanied by a charged lepton with 
$\Pl>1$~\gevc, through $N_\mathrm{sl}^\mathrm{true}=(N_\mathrm{sl}-BG_\mathrm{sl})/ \epsilon_\ell^\mathrm{sl} \epsilon^\mathrm{sl}_\mathrm{tag}$. 
Here, $BG_\mathrm{sl}$ is the remaining
peaking background estimated from MC simulation and $N_\mathrm{sl}$ is obtained from 
the \mes\ fit to the selected semileptonic sample and 
$\epsilon_\ell^\mathrm{sl}$ refers to the efficiency for selecting a lepton from a 
semileptonic $B$ decay with a momentum $\Pl>1$~\gevc in an event tagged with 
efficiency $\epsilon^\mathrm{sl}_\mathrm{tag}$.
We obtain $N_\mathrm{sl} = 237,433 \pm 838$ 
and $BG_\mathrm{sl} = 20,705 \pm 132$.
 
The ratio of efficiencies in Eq.~(\ref{eq:ratioBR})
accounts for differences in the final states and the different lepton momentum spectra for the
two classes of events, and their impact on the tagging. 
The efficiencies for \breco tagging and lepton detection are not very different, and thus the efficiency ratio is close to one. 

We convert Eq.~(\ref{eq:ratioBR}) to partial branching fractions
by using the total semileptonic branching fraction, 
${\cal B}(\Bxlnu)=(10.75\pm0.15)\%$~\cite{PDG2010}.

The regions of phase space, fitted event yields, efficiencies introduced in Eq.~(\ref{eq:ratioBR}), 
and partial branching fractions are listed in Table~\ref{tab:inputdeltaB};
the regions are one-dimensional in \mX, \Pplus, or \Pl, 
or two-dimensional in the plane \mX versus \Q. In the following, we will
refer to the latter as \mX\ -- \Q.
Two fits have been performed with no additional kinematic restrictions, apart
from the requirement $\Pl > 1$~\gevc:
a fit to the lepton momentum spectrum and a fit to the two-dimensional 
histogram \mX\ -- \Q.  Since the same events enter both fits, the
correlation is very high. The fact that the results are in excellent 
agreement indicates that
the distribution of the simulated signal and background distributions 
agree well with the data.

Correlations between the different analyses are reported in the entries above the main diagonal of 
Table~\ref{tab:statcorr}.

In addition, a series of fits to the lepton momentum spectrum has been performed  with the lower limit
on \Pl increasing from 1.0~\gevc to 2.4~\gevc.
The results are presented in Section~\ref{sec:theonew};
the measurement at $\Pl > 1.3$~\gevc gives the smallest total uncertainty and is also quoted 
in Table~\ref{tab:inputdeltaB}.

Consistency checks have been performed. The analysis done on data samples
collected in different data-taking periods, or separating the lepton flavor or charge have all 
yielded the same results, within experimental uncertainties.

\begin{figure}[htb]
\includegraphics[width=0.33\textwidth]{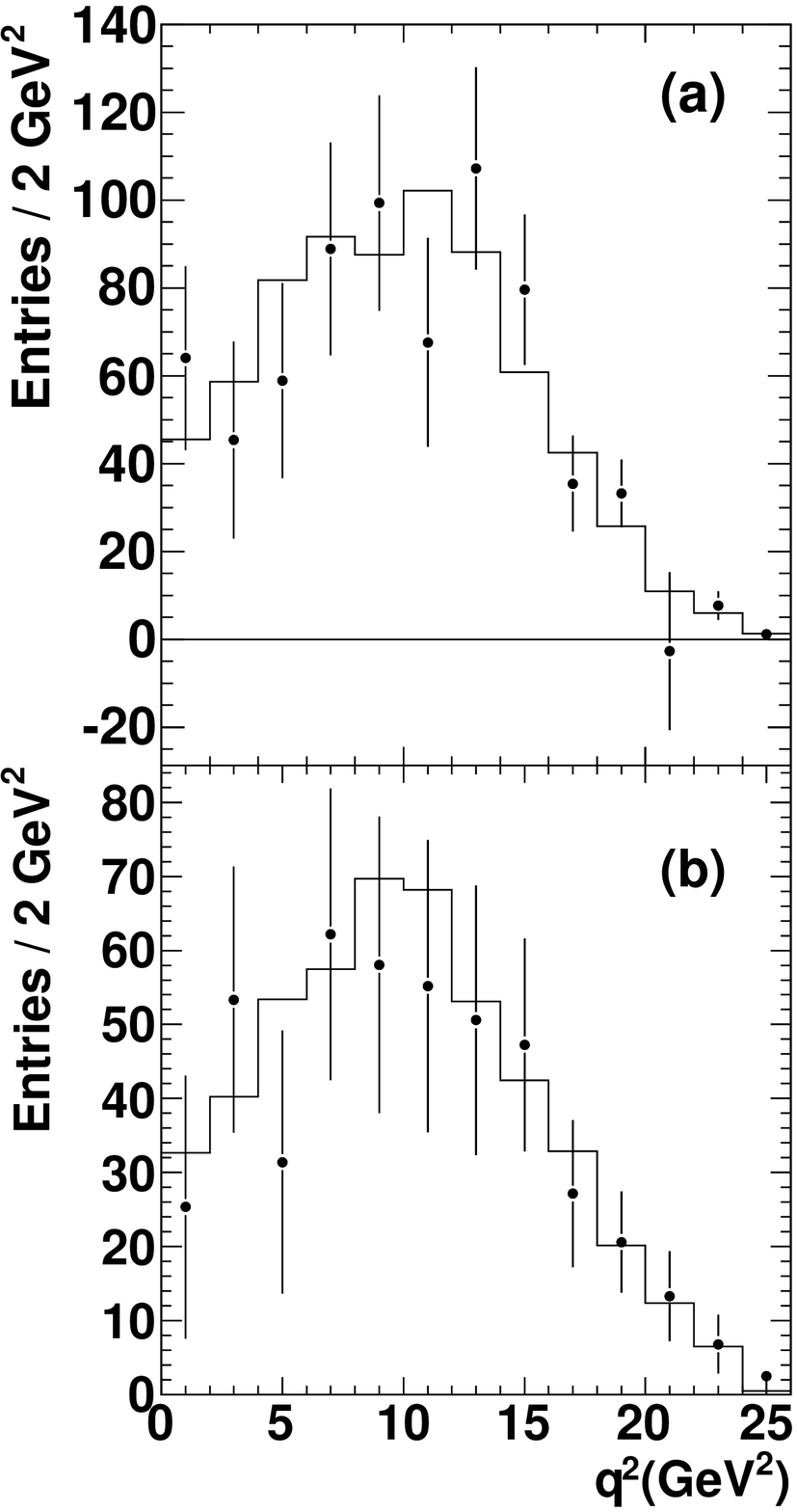}\\
\caption{Comparison of the measured \Q distributions (data points) for $\mX<1.7$~\gevcc for charmless 
semileptonic decays of (a) charged  and (b) neutral 
$B$ mesons to the results of the fit (histogram),  after \Bxclnu and ``other'' background subtraction.}
\label{fig:mxq2separated}
\end{figure}

\begin{table*}[htb]
\begin{center}
\caption{List of the fitted numbers of signal events $N_u$, the number of 
         events generated outside the kinematic selection $N_u^\mathrm{out}$, the efficiencies, 
         the partial branching fractions $\Delta \BR(\Bxulnu)$ and the $\chi^2$ per degree of freedom 
         for the different selected regions of phase space. 
         The first uncertainty is statistical, the second systematic.
         The $\Pl>1$~\gevc requirement is implicitly assumed. 
}
\vspace{0.1in}
\begin{tabular*}{\textwidth}{l@{\extracolsep{\fill}}cccccc} 
\hline\hline
Region of phase space   & $N_u$ & $N_u^\mathrm{out}$  & $\epsilon_\mathrm{sel}^u \epsilon_\mathrm{kin}^u$ &${(\epsilon_\ell^\mathrm{sl} \epsilon_t^\mathrm{sl})} /{(\epsilon_\ell^u \epsilon_t^u )}$  & $\Delta \BR(\Bxulnu)\ (10^{-3})$ & $\chi^2$/ndof\\ \hline
$\mX < 1.55$~\gevcc  	& $1033 \pm 73$	 & $29 \pm 2$ 	& $0.365\pm0.002$ & $1.29\pm0.03$ &  $1.08 \pm 0.08 \pm 0.06$ & 7.9/8 \\
$\mX < 1.70$~\gevcc  	& $1089 \pm 82$  & $25 \pm 2$ 	& $0.370\pm0.002$ & $1.27\pm0.04$ &  $1.15 \pm 0.10 \pm 0.08$ & 6.6/8 \\
$\Pplus < 0.66$~\gevc	    & $ 902 \pm 80$  & $54 \pm 5$ 	& $0.375\pm0.003$ & $1.22\pm0.03$ &  $0.98 \pm 0.09 \pm 0.08$ & 3.4/9 \\
$\mX < 1.70$~\gevcc, $\Q > 8$~\gevccsq	& $ 665 \pm 53$	 & $39 \pm 3$  	& $0.386\pm0.003$ & $1.25\pm0.03$ &  $0.68 \pm 0.06 \pm 0.04$ & 23.7/26\\
\mX\ -- \Q                & $ 1441 \pm102$& $ 0 $  & $0.338\pm0.002$ & $1.18\pm0.03$ &  $1.80 \pm 0.13 \pm 0.15$ & 31.0/29 \\
$\Pl>1.0$~\gevc          & $1470 \pm  130$  & $8 \pm 2$ & $0.342 \pm 0.002$ & $1.18 \pm0.03$ & $1.81 \pm 0.16 \pm 0.19$ & 21.6/14 \\
$\Pl>1.3$~\gevc          & $1329 \pm  121$  & $61 \pm 5$ & $0.363\pm0.002$ & $1.18 \pm0.09$ & $1.53 \pm 0.13 \pm 0.14$  & 20.4/14 \\
\hline\hline
\end{tabular*}
\label{tab:inputdeltaB}
\end{center}
\end{table*}

\begin{table*}[htb]
\begin{center}
\caption{Correlation coefficients for measurements in different kinematic 
regions. The entries above the main diagonal refer to correlations (statistical and systematic)
for pairs of measurements of the partial branching fractions; the entries below the diagonal refer to the 
correlations (experimental and theoretical)
for pairs of \Vub measurements.}
\vspace{0.1in}
\begin{tabular*}{\textwidth}{l@{\extracolsep{\fill}}cccccc}
\hline\hline\vspace{1mm}
\multirow{2}{35mm}{Phase space restriction} & $\mX < 1.55$	& $\mX < 1.70$ 	& $\Pplus < 0.66$ & $\mx < 1.70\gevcc,$ 	& \mX\ -- \Q & $\Pl>1.3$ \\
	 &	\gevcc	&	\gevcc	&	\gevc	  & $\Q > 8$~\gevccsq	& $\Pl>1.0$~\gevc & \gevc  \\   
\hline
$\mX < 1.55$~\gevcc  				&   1		&      0.77 	&      0.74	&      0.50	&      0.72	&  0.57	\\
$\mX < 1.70$~\gevcc  				& 0.81	&   	1	&      0.86 	&      0.55	&      0.94	&  0.73	\\
$\Pplus < 0.66$~\gevc				& 0.69	&  0.81	&   	1	&      0.46	&      0.78	&  0.61	\\
$\mX < 1.70$~\gevcc, $\Q > 8$~\gevccsq	& 0.40	&  0.46	&  0.38	&   	1	&      0.52	&  0.46	\\
\mX\ -- \Q   		         	& 0.58	&  0.88	& 0.67	&  0.34	&       1	&  0.74	\\
$\Pl>1.3$~\gevc 				& 0.53	&  0.72	&  0.58	&  0.40	&  0.72   	&   1	\\
\hline\hline
\end{tabular*}
\label{tab:statcorr}
\end{center}
\end{table*}

\subsection{Partial branching fractions for \Bzb and \Bm}
All the fits, except those to the \Pl distribution, have been repeated
separately for charged and neutral \breco tags. 
In this case, we extract the true signal yields from 
the measurements by the following relations to determine the partial branching fractions:
\begin{eqnarray*}
N^0_\mathrm{meas}&=&{\cal P}_{\Bzb_\mathrm{true}\to \Bzb_\mathrm{reco}} N^0_\mathrm{true}+{\cal P}_{\Bm_\mathrm{true}\to \Bzb_\mathrm{reco}} N^{-}_\mathrm{true}, \\
N^{-}_\mathrm{meas}&=&{\cal P}_{\Bzb_\mathrm{true}\to \Bm_\mathrm{reco}} N^0_\mathrm{true}+{\cal P}_{\Bm_\mathrm{true}\to \Bm_\mathrm{reco}} N^{-}_\mathrm{true},
\label{eq:crossfeed}
\end{eqnarray*}
where the cross-feeds probabilities,
${\cal P}_{\Bm_\mathrm{true}\to \Bzb_\mathrm{reco}}$ 
and 
${\cal P}_{\Bzb_\mathrm{true}\to \Bm_\mathrm{reco}}$, 
are computed using MC simulated events and are typically of the order of (2 - 3)\%.

Figure~\ref{fig:mxq2separated} shows the \Q distributions of \Bxulnu
events after background subtraction, for charged and neutral \B decays, with $\mX<1.7$~\gevcc. 
Fitted yields, efficiencies, and partial branching fractions are given in Table~\ref{tab:inputdeltaB0Bp}. 

\begin{table*}[htb]
\begin{center}
\caption{Summary of the fits  to separate samples of neutral and charged \B decays. For details see Table~\ref{tab:inputdeltaB}.}
\vspace{0.1in}
\begin{tabular*}{\textwidth}{l@{\extracolsep{\fill}}cccccc} 
\hline\hline\vspace{1mm}
\rule{0pt}{9pt}\Bzb decays & $N_u$ & $N_u^\mathrm{out}$  & $\epsilon_\mathrm{sel}^u \epsilon_\mathrm{kin}^u$ &${(\epsilon_\ell^\mathrm{sl} \epsilon_t^\mathrm{sl})} /{(\epsilon_\ell^u \epsilon_t^u )}$  & $\Delta \BR(\Bxulnu)\ (10^{-3})$ & $\chi^2$/ndof\\ \hline
$\mX < 1.55$~\gevcc  				& $458 \pm 48$	 & $12 \pm 1$ 	& $0.360\pm0.004$ & $1.49\pm0.07$ &  $1.09 \pm 0.12 \pm 0.11$ & 19.0/9 \\
$\mX < 1.70$~\gevcc  				& $444 \pm 53$  & $12 \pm 1$ 	& $0.370\pm0.004$ & $1.45\pm0.07$ &  $1.12 \pm 0.11 \pm 0.11$ & 16.6/9 \\
$\Pplus < 0.66$~\gevc				& $434 \pm 52$  & $27 \pm 3$ 	& $0.367\pm0.004$ & $1.38\pm0.06$ &  $1.09 \pm 0.13 \pm 0.11$ & 9.1/9 \\
$\mX < 1.70$~\gevcc, $\Q > 8$~\gevccsq	& $262 \pm 38$	 & $16 \pm 2$  	& $0.380\pm0.005$ & $1.43\pm0.06$ &  $0.61 \pm 0.09 \pm 0.06$  & 15.8/26 \\
\mX\ -- \Q                            & $ 553 \pm 72$& $ 0 $  & $0.328\pm0.003$ & $1.36\pm0.08$ &  $1.58 \pm 0.21 \pm 0.20  $ & 14.8/29\\
\hline
\vspace{1mm}
\rule{0pt}{9pt}\Bm decays & $N_u$ & $N_u^\mathrm{out}$  & $\epsilon_\mathrm{sel}^u \epsilon_\mathrm{kin}^u$ &${(\epsilon_\ell^\mathrm{sl} \epsilon_t^\mathrm{sl})} /{(\epsilon_\ell^u \epsilon_t^u )}$  & $\Delta \BR(\Bxulnu)\ (10^{-3})$ & $\chi^2$/ndof \\ \hline
$\mX < 1.55$~\gevcc  				& $591 \pm 56$	 & $17 \pm 2$ 	& $0.370\pm0.003$ & $1.18\pm0.04$ &  $1.12 \pm 0.11 \pm 0.11$ & 3.1/9\\
$\mX < 1.70$~\gevcc  				& $669 \pm 63$  & $14 \pm 1$ 	& $0.370\pm0.003$ & $1.17\pm0.07$ &  $1.27 \pm 0.14 \pm 0.13$ & 3.3/9 \\
$\Pplus < 0.66$~\gevc				& $ 491 \pm 61$  & $28 \pm 4$ 	& $0.379\pm0.004$ & $1.11\pm0.03$ &  $0.96 \pm 0.12 \pm 0.12$ & 2.0/9 \\
$\mX < 1.70$~\gevcc, $\Q > 8$~\gevccsq	& $ 406 \pm 41$	 & $24 \pm 2$  	& $0.392\pm0.004$ & $1.43\pm0.03$ &  $0.74 \pm 0.08 \pm 0.08$ & 26.9/26 \\
\mX\ -- \Q                            & $ 859 \pm79$& $ 0 $  & $0.345\pm0.003$ & $1.07\pm0.03$ &  $1.91 \pm 0.18 \pm 0.22  $ & 36.7/29\\
\hline\hline
\end{tabular*}
\label{tab:inputdeltaB0Bp}
\end{center}
\end{table*}

\subsection{Data - Monte Carlo comparisons}
\label{sec:thecomparison}
The separation of the signal events from the noncombinatorial backgrounds
relies heavily on the MC simulation to correctly describe the distribution for signal and 
background sources. Therefore, an extensive study has been devoted to detailed comparisons of data 
and MC distributions. 

\begin{figure*}[htb]
 \begin{centering}
 \includegraphics[width=0.34\textwidth]{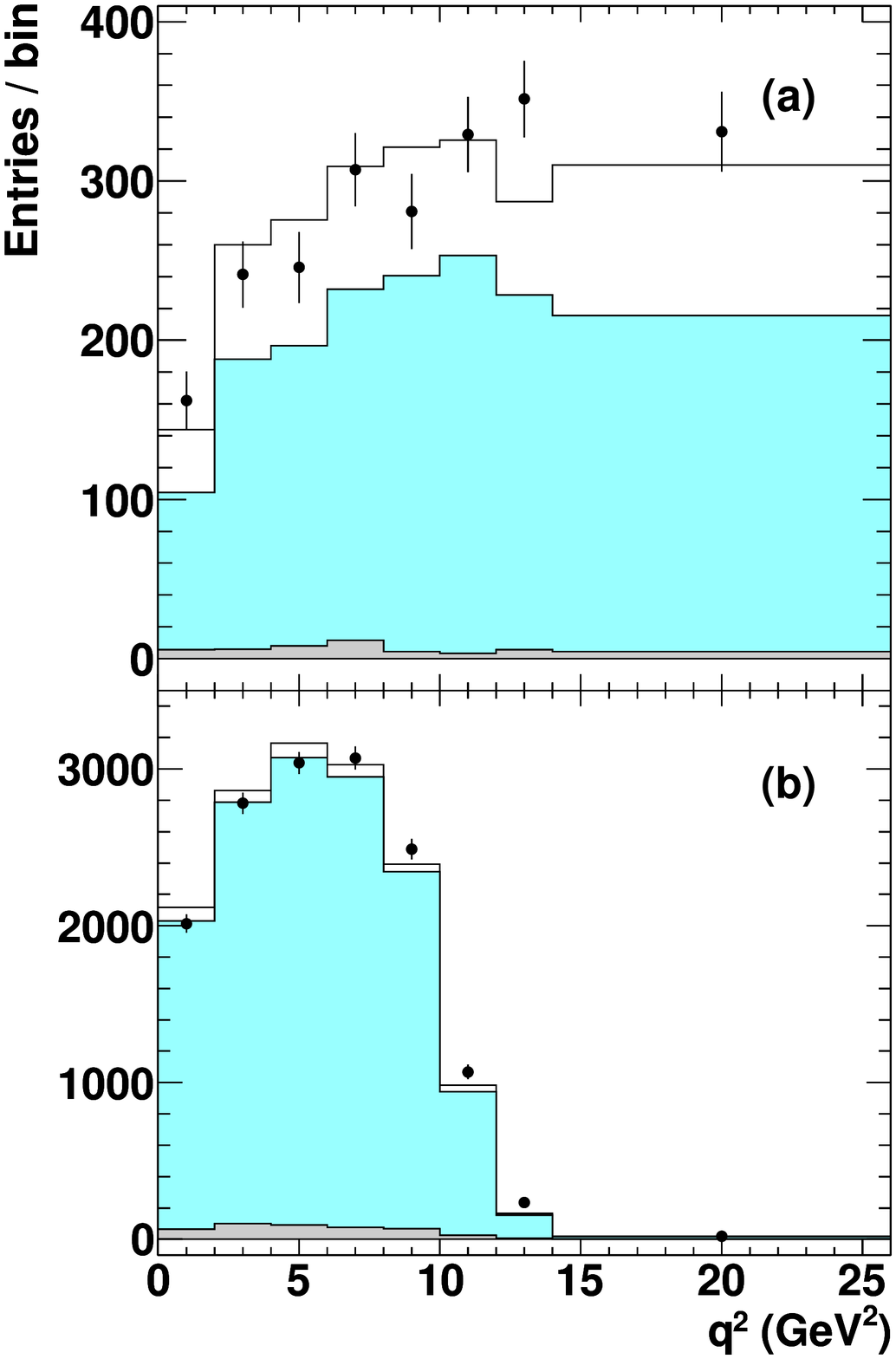} 
 \includegraphics[width=0.34\textwidth]{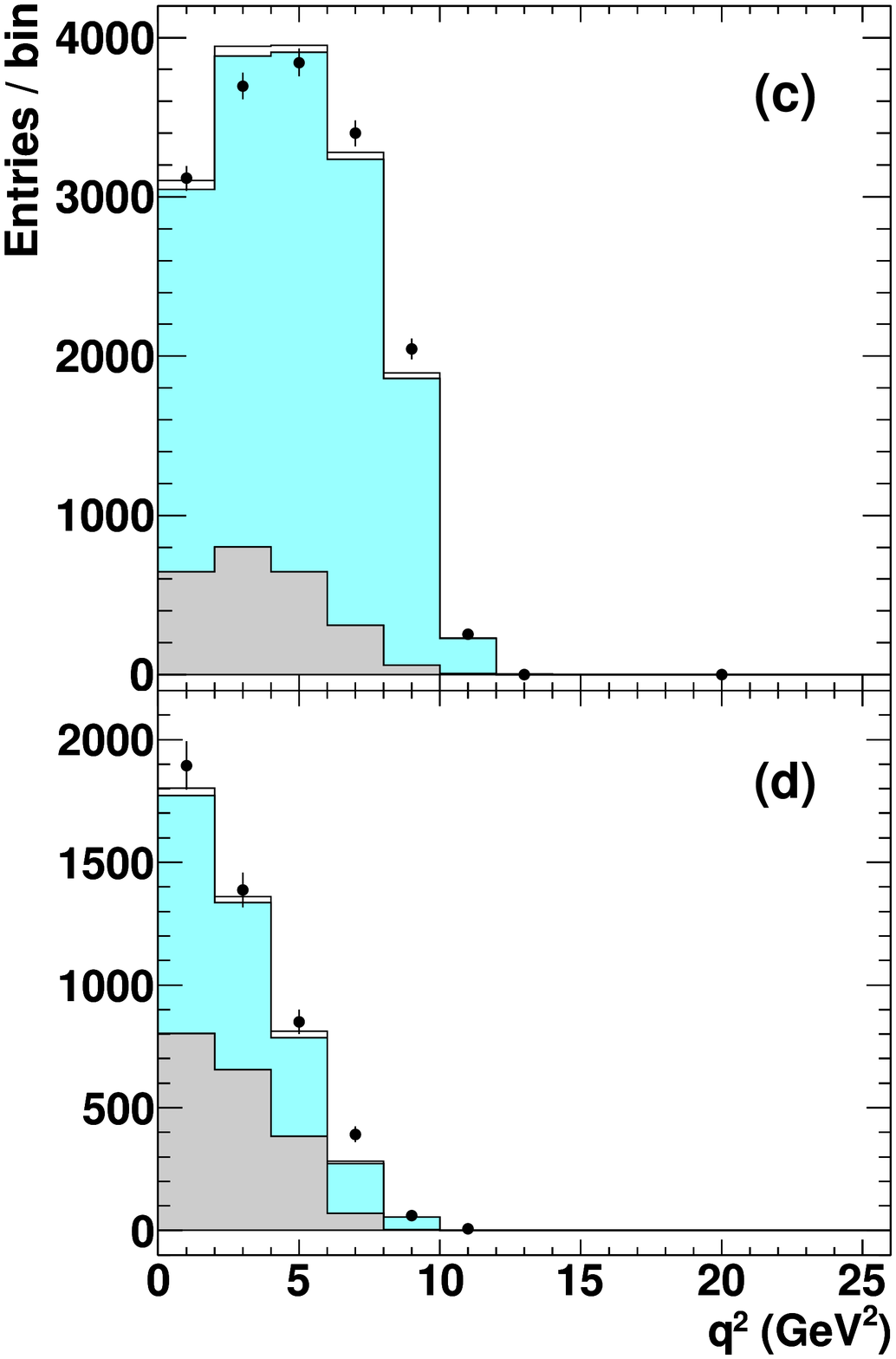} 
\caption{Fit results to the \mX\ -- \Q distribution for the signal-depleted sample. 
The \Q\ distribution is reported separately for the four \mX bins:
(a) $\mX\le 1.5$~\gevcc,  (b) $1.5<\mX\le 2.0$~\gevcc,  (c) $2.0<\mX\le 2.5$~\gevcc and
(d) $2.5<\mX\le 3.0$~\gevcc. The three MC contributions shown here are: \Bxulnu decays vetoed by 
the selection (no shading), \Bdlnu, \Bdslnu and ``other'' background 
(light shading), and the \Bdsslnu component as defined in the text (dark shading).  
\label{fig:lambdadss}}
 \end{centering}
\end{figure*}

A correction applied to the simulation improves the quality of the fits to the 
kinematic distributions in regions that are dominated by \Bxclnu background, 
especially in the high \mx region.
In the simulation, we adjust $\lambda_{D^{**}}$, the ratio of branching fractions of semileptonic 
decays to $P$-wave $D$ mesons and nonresonant charm states decaying to $D^{(*)}X$, 
over the sum of all $D^{(*)} \ell \bar{\nu}$ and ``other'' background components,
\begin{equation}
\label{eq:lambda}
\lambda_{D^{**}} = \frac
	{\BR(\Bdsslnu) + \BR(\Bdbsxlnu)}
	{\BR(\Bdbslnu)+\BR(\Bbar\rightarrow X_\mathrm{other})}.\\
\end{equation}
This ratio has been determined from data by performing a fit 
on the \mX\ -- \Q distribution of the signal-depleted sample without kinematic selection. 
The resulting distribution of this fit is shown in Fig.~\ref{fig:lambdadss}.
We measure $\lambda_{D^{**}} = 0.73 \pm 0.08$, where the error 
takes into account the fact that $\chi^2/\mathrm{ndof} = 2$.
Other determinations, using signal-enriched samples, give statistically consistent results. 
This adjustment improves the quality of the fits in regions where backgrounds dominate, 
but it has a small impact on the fitted signal yield.
We have verified that using $D^{**}$ MC correction factors determined separately on each analysis 
do not change significantly the results with respect to 
our default strategy, where $\lambda_{D^{**}}$ is determined for the most 
inclusive sample available, namely the signal-depleted sample 
of the analysis without kinematic requirements.

\begin{figure*}[htb]
\includegraphics[width=\textwidth]{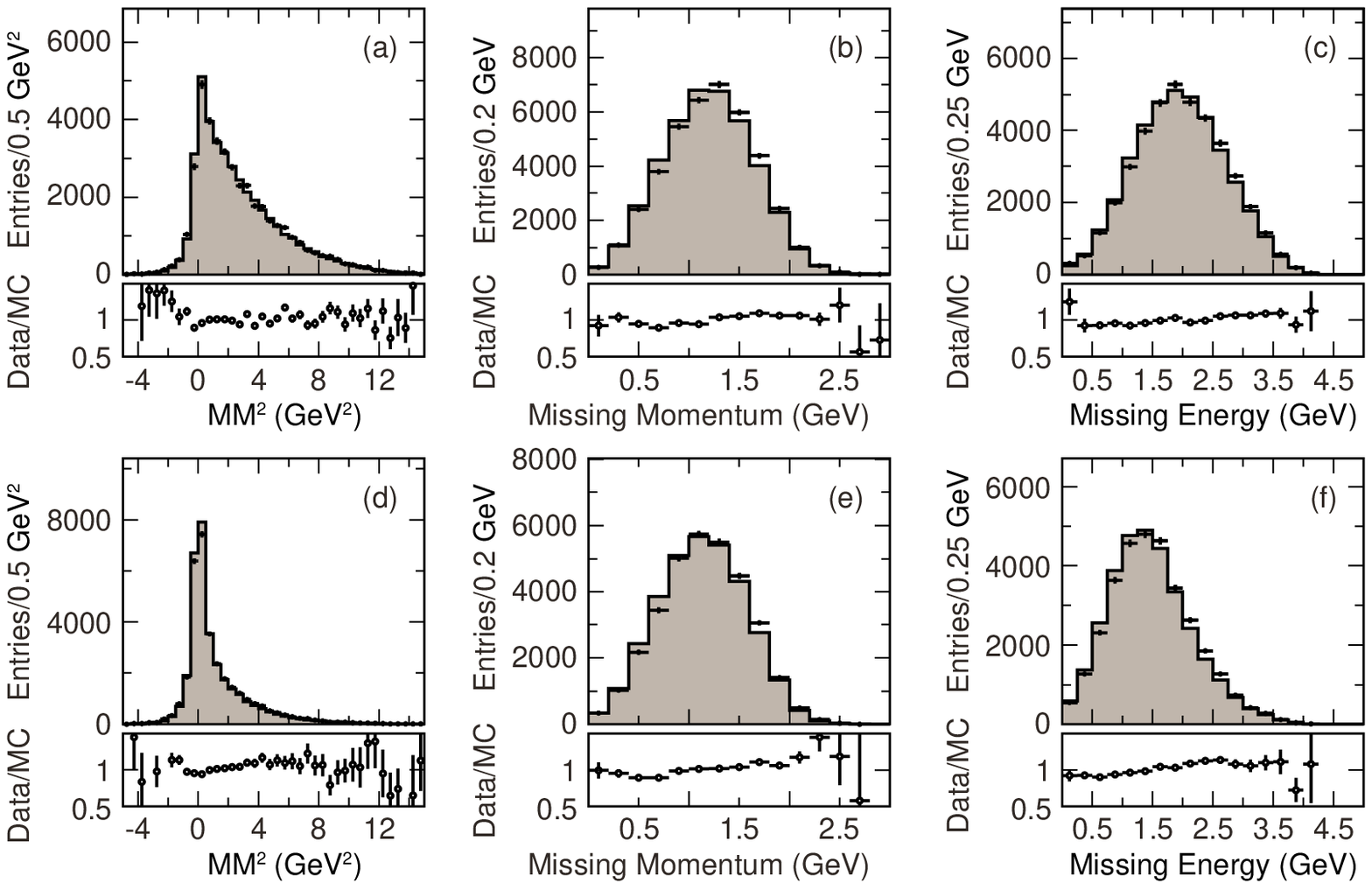}
\caption{Comparison of data (points with statistical uncertainties) and MC (histograms) simulated 
distributions of (a,d) the missing mass squared, (b,e) the missing momentum, and (c,f)  
the missing energy  for \Bxulnu enhanced (top row) and  depleted  (bottom row) event samples.}
\label{fig:theComp:mm2}
\end{figure*}

\begin{figure*}[htb]
\includegraphics[width=\textwidth]{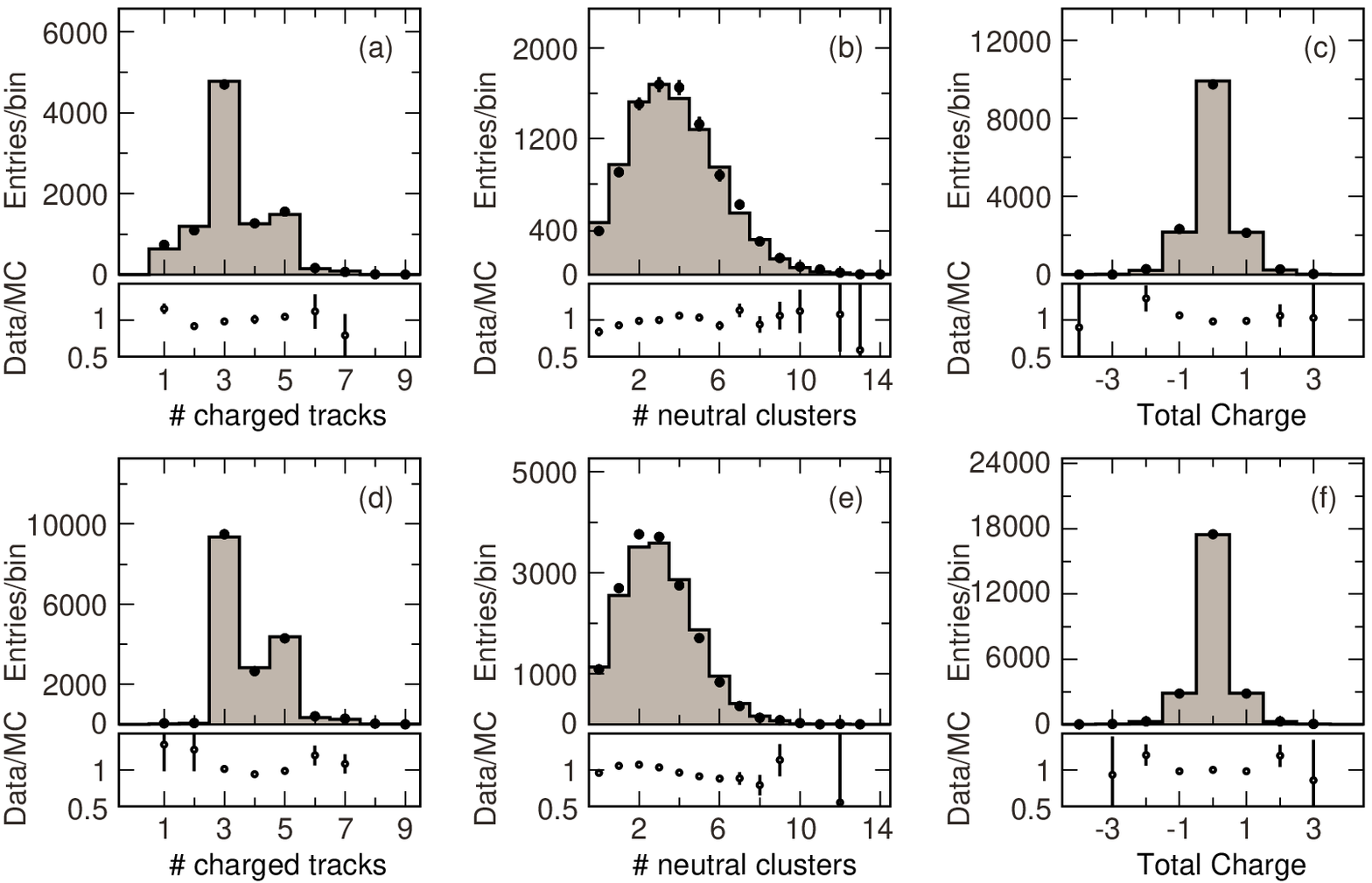}
\caption{ 
Comparison of data (points with statistical uncertainties) and MC (histograms) simulated distributions of 
(a,d) the charged track multiplicity, (b,e) the photon multiplicity, (c,f) and the total charge per 
event for \Bxulnu enhanced (top  row) and  depleted  (bottom row) event samples.}
\label{fig:theComp:nchg}
\end{figure*}

Figures~\ref{fig:theComp:mm2} and~\ref{fig:theComp:nchg} show 
comparisons of data and MC distributions, after subtraction of the combinatorial background,
for signal-enriched and signal-depleted event samples.
All the selection criteria have been applied, except those affecting directly the variable shown.  
The spectra are background-subtracted based on the results of the \mes fit
performed for each bin of the variable shown. The uncertainties on data points 
are on the yields of the bin-by-bin fits. The data and MC distributions are normalized to the same area. 
The overall agreement is reasonable, taking into account that the uncertainties are purely statistical. 
The effects that introduce differences between data and simulation are described in 
Section~\ref{sec:systematics}; their impact is assessed and accounted for as systematic uncertainty.

\section{Systematic uncertainties}
\label{sec:systematics}
The experimental technique described in this article, namely the measurement of a ratio of
branching fractions, ensures that systematic uncertainties due, for example, to radiative
corrections or differences between \Bpm and \Bz or \Bzb production rate and lifetime, are negligible.
A summary of all other statistical and systematic uncertainties on the partial
branching fractions for selected kinematic regions of phase space is
shown in Table~\ref{tab:systematics} for the complete data sample,
and in Table~\ref{tab:sysall:ChargeSep} for charged and neutral \B samples
separately. 

The individual sources of systematic uncertainties are, to a good
approximation, uncorrelated and can therefore be added in quadrature
to obtain the total systematic uncertainties for partial branching fraction.  
In the following, we discuss the assessment of the systematic uncertainties in detail. 

To estimate the systematic uncertainties on the ratio $\Delta R_{u/sl}$, we
compare the results obtained from the nominal fits with results
obtained after changes to the MC simulation that reflect the
uncertainty in the parameters which impact the detector efficiency and
resolution or the simulation of signal and background processes.  
For instance, we lower the tracking efficiency by randomly eliminating a
fraction of tracks (corresponding to the estimated uncertainty) in the MC sample, 
redo the event reconstruction and selection on the recoil side,
perform the fit, and take the
difference compared to the results obtained with the nominal MC
simulation as an estimate of the systematic uncertainty.  The sources of
systematic uncertainties are largely identical for all selected signal
samples, but the size of their impact varies slightly.

 \begin{table*}[htb]
   \begin{center}
     \caption{\small Statistical and systematic uncertainties (in percent) on measurements of the 
     partial branching fraction in seven selected kinematic regions. The total systematic 
     uncertainty is the sum in quadrature of the MC statistical uncertainty and all the other 
     single contibutions from detector effects, signal and background simulation, background 
     subtraction and normalization. The total uncertainty is the sum in quadrature of the data 
     statistical and total systematic uncertainties.} 
    \vspace{0.1in}
\begin{tabular}{lc@{\extracolsep{8pt}}c@{\extracolsep{8pt}}c@{\extracolsep{8pt}}c@{\extracolsep{8pt}}c@{\extracolsep{8pt}}c@{\extracolsep{8pt}}c} 
\hline\hline
\multirow{2}{35mm}{Phase space restriction} 	&  $\mX < 1.55$	&  $\mX < 1.70$ 	&  $\Pplus < 0.66$ 	&  $\mX < 1.70\gevcc,$ &  \multirow{2}{15mm}{\mX\ -- \Q} & $\Pl>1.0$ & $\Pl>1.3$ \\ 
                                            	&  \gevcc    	&  \gevcc	        &  \gevc         	&  $\Q > 8$~\gevccsq   &                                   & \gevc     & \gevc      \\   
\hline       
 Data statistical uncertainty     &  7.1           & 8.9 & 8.9           &  8.0  	&  7.1	&  9.4 & 8.8	\\
 MC statistical uncertainty &  1.3           & 1.3 & 1.3           &  1.6  	&  1.1	&  1.1 & 1.2	\\ \hline
 \multicolumn{8}{c}{Detector effects } \\ \hline
 Track efficiency           & 0.4  & 1.0 & 1.1 & 1.7 & 0.7 & 1.2 & 1.0	\\         
 Photon efficiency          & 1.3  & 2.1 & 4.0 & 0.7 & 1.0 & 0.9 & 0.9	\\                                                                                                                                   
 \piz	 efficiency         & 1.2 & 0.9 & 1.1 & 0.9 & 0.9 & 2.9 & 1.1 \\
 Particle identification    & 1.9 & 2.4 & 3.3 & 2.9 & 2.3 & 2.9 & 2.2 \\    
$K_L$ production/detection  & 0.9 & 1.3 & 1.1 & 2.1 & 1.6 &	1.3 & 0.6 \\
$K_S$ production/detection  &  	0.8     & 1.4   &  	1.7  &  	2.1  &  	1.2  & 1.3 & 0.3  \\ \hline     
 \multicolumn{8}{c}{Signal simulation  } \\\hline 
Shape function parameters  &  2.0 & 1.3 & 1.2 & 0.7 & 5.4 & 6.4 & 6.6 \\
Shape function form                      & 	1.2     & 1.6&	2.6  & 		1.2  &  	1.5   & 1.1 & 1.1 \\
 Exclusive \Bxulnu             &  	0.6	& 1.3&	1.6  &  	0.7  &  	1.9   & 5.3 &  3.4\\
\ssbar production               & 	1.2  	& 1.6&	1.1  & 		1.0  &  	2.7  & 3.1 & 2.4   \\ \hline      
 \multicolumn{8}{c}{Background simulation   } \\ \hline
 $B$ semileptonic branching ratio       &  	0.9     & 1.4   &  	1.5  &  	1.4  &  	1.0  & 0.8 &  0.7  \\   
 $D$ decays                   & 	1.1     & 0.6   &	1.1  & 		0.6  &  	1.1  & 1.6 & 1.5  \\      
$\B\to\D\ell\nu$ form factor  		& 0.5   & 0.5   & 1.3 & 0.4 & 0.4 & 0.1 & 0.2                       \\
$\B\to\Dstar\ell\nu$ form factor  	& 0.7   & 0.7   & 0.9 & 0.7 & 0.7 & 0.7 & 0.7                        \\
$\B\to D^{**}\ell\nu$ form factor  	& 0.8   & 0.9   & 1.3 & 0.4 & 0.9 & 1.0 & 0.3                        \\ 
$\B\to D^{**}$ reweighting          & 0.5   & 1.4   & 1.5 & 1.0 & 1.9 & 0.4 & 1.5                        \\ \hline                                       
\multicolumn{8}{c} {\mes\ background subtraction}       \\  \hline      
 $m_{ES}$ background subtraction      & 	2.0    	& 2.7 & 	1.9  &  	2.6  & 		1.9 & 2.0 & 2.5	\\   
 combinatorial backg.         &  	1.8  	& 1.8 &	2.6  &  	1.8  & 		1.0 & 2.1 & 0.5	\\ \hline
 \multicolumn{8}{c} {Normalization}       \\  \hline
 Total semileptonic BF &  1.4   & 1.4   & 1.4   & 1.4   & 1.4   & 1.4   & 1.4 \\ \hline
Total systematic uncertainty       & 5.5 & 6.7 & 8.3 & 6.6 & 8.4 & 11.0 & 9.3 \\
Total experimental uncertainty     & 9.0 & 11.1 & 12.2 & 10.4 & 11.0 & 14.4 & 12.8 \\  \hline\hline
\end{tabular}
\label{tab:systematics}
\end{center}
\end{table*}

 \begin{table*}[htb]
   \begin{center}
     \caption{\small Statistical and systematic uncertainties (in percent) on the partial branching 
     fraction for neutral and charged \B mesons for the five selected kinematic regions. 
     The total systematic uncertainty is the sum in quadrature of the MC statistical uncertainty 
     and all the other single contibutions from detector effects, signal and background simulation, 
     background subtraction and normalization. The total uncertainty is the sum in quadrature of 
     the the data statistical and total systematic uncertainties.}
     \vspace{0.1in}
\newcolumntype{Y}{>{\centering\arraybackslash}X}
 \begin{tabularx}{0.96\textwidth}{>{\setlength{\hsize}{3.3\hsize}}X>{\setlength{\hsize}{0.75\hsize}}Y>{\setlength{\hsize}{0.75\hsize}}Y>{\setlength{\hsize}{0.75\hsize}}Y>{\setlength{\hsize}{0.75\hsize}}Y>{\setlength{\hsize}{0.75\hsize}}Y>{\setlength{\hsize}{0.75\hsize}}Y>{\setlength{\hsize}{0.85\hsize}}Y>{\setlength{\hsize}{0.85\hsize}}Y>{\setlength{\hsize}{0.75\hsize}}Y>{\setlength{\hsize}{0.75\hsize}}Y}
 \hline\hline
\multirow{2}{35mm}{Phase space restriction}       &  \multicolumn{2}{c}{$\mX < 1.55$}	&  \multicolumn{2}{c}{$\mX < 1.70$} 	&  \multicolumn{2}{c}{$\Pplus < 0.66$} 	&  \multicolumn{2}{c}{$\mX < 1.70\gevcc,$}        &  \multicolumn{2}{c}{    }\\
   	  &  \multicolumn{2}{c}{\gevcc}  	    &  \multicolumn{2}{c}{\gevcc}	        &  \multicolumn{2}{c}{\gevc}         	&  \multicolumn{2}{c}{$\Q > 8$~\gevccsq}                         &  \multicolumn{2}{c}{\raisebox{1.5ex}[0cm][0cm]{\mX\ -- \Q}}  \\ 
      & \Bzb & \Bm & \Bzb & \Bm & \Bzb & \Bm & \Bzb & \Bm & \Bzb & \Bm \\ 
\hline 
%%%%%%%%%%%%%%%%%%%%%%%%%%%%%%%%%            MX1.55         Mx1.7           PPLUS        MXQ2          FULL 
       Data statistical uncertainty      		&  \parbox{0.9cm}{10.4}  & \parbox{0.9cm}{9.6} &  \parbox{0.9cm}{14.4} & \parbox{0.9cm}{11.0}	&  \parbox{0.9cm}{12.0} & \parbox{0.9cm}{12.5}	&  14.6 & 10.1 &  \parbox{0.9cm}{13.0} & \parbox{0.9cm}{9.2} \\
       MC statistical uncertainty 		      &   2.5  & 1.6 &   2.5 &  1.6	&   2.4 & 1.8 	&   2.8 &  2.0 &  1.9  & 1.3 \\ 
 Detector effects  		    	&   4.5  & 4.9 &   5.0 &  6.3	&   5.9 & 7.2   &   5.3 &  6.3 &  5.6  & 4.7 \\         
 Signal simulation & 6.6 & 5.4 & 5.2 & 4.8 & 4.8 & 5.9 & 3.7 & 5.4 & 8.7 & 7.6 \\
 Background simulation                    &   4.4  & 4.2 & 5.6   & 4.5   & 5.9   & 5.4   & 4.9   & 4.4	& 4.4   & 5.0 \\
\mes\ background subtraction    &   4.1  & 5.4 &   5.2  & 5.0	&   2.9 & 5.3	&   5.2 & 5.1  &  3.8  & 4.1 \\    
 Total semileptonic BF  &  1.4   & 1.4   & 1.4   & 1.4   & 1.4   & 1.4   & 1.4 & 1.4   & 1.4   & 1.4 \\
 %%%%%%%%%%%%%%%%%%%%%%%%%%%%%%%%%%%%%%%%%%%%%%%%%%%%%%%%%%%%%%%%%%%%%%%%%%%%%%%%%%%%%%%%%%%%%%%%%%%%%%%%%%%%%%%%%%%%%%%%%%%%%%%%%%%%%%%%%%%%%%%%%%%%%%
 \hline
Total systematic uncertainty  & 10.4 & 10.2 & 10.9 & 10.6 & 10.5 & 12.2 & 10.1 & 11.0 & 12.1 & 11.2 \\
Total experimental uncertainty & 14.7 & 14.0 & 18.1 & 15.3 & 15.9 & 17.5 & 17.8 & 14.9 & 17.8 & 14.5 \\
\hline\hline 
\end{tabularx}     
     \label{tab:sysall:ChargeSep}
   \end{center}
 \end{table*}

\subsection{Detector effects}
Uncertainties in the reconstruction efficiencies for charged and neutral particles, in the rate of 
tracks and photons from beam background, misreconstructed tracks, failures in the matching of EMC clusters to 
charged tracks, showers split-off from hadronic interactions, undetected $K_L$, and additional 
neutrinos, all contribute to the event reconstruction and impact the variables that are used in 
the event selection and the analysis. For all these effects the uncertainties in the efficiencies 
and resolution have been derived from comparisons of data and MC simulation for selected control samples.

From the study of the angular and momentum distributions of low momentum pions in $D^{*}$ samples, 
we estimate the uncertainty on the track finding efficiency at low momenta to be about $1.0\%$.
For all other tracks, the difference between data and MC in tracking efficiency is estimated to be 
about~$0.5\%$ per track. The systematic uncertainty on the ratio $\Delta R_{u/sl}$ is calculated
as described above, and shown in Tables~\ref{tab:systematics} and~\ref{tab:sysall:ChargeSep}.

Similarly, for single photons, we estimate the systematic uncertainty by randomly eliminating showers 
that are not matched to the $\pi^0_\mathrm{soft}$ used to veto $\Bdslnu$ decays,
with a probability of 1.8\% per shower.

We estimate the systematic uncertainty due to \piz detection by randomly eliminating neutral
pions that are used in the \Bdslnu veto, with a probability of 3\% per \piz.

Uncertainties on charged particle identification efficiencies have been assessed to be
2.0\% for electrons and 3.0\% for muons. The uncertainty on the 
corresponding misidentification rates are estimated to be 15\%.
Systematic uncertainties on the kaon identification efficiency and misidentification rate are 
2\% and 15\%, respectively.

In this analysis, no effort was made to identify \KL.
On the other hand, \KL mesons interacting in the detector deposit only a fraction of their energy in the 
EMC, thus they impact $P_\mathrm{miss}$ and other kinematic variables used in this analysis.
Based on  detailed studies of data control samples of $D^0 \to K^0 \pi^+\pi^-$ decays, 
corrections to the \KL efficiency and energy deposition have been 
derived and applied to the simulation as a function of the $K^0_L$ momentum and angles. 
We take the difference compared to the results obtained without this correction applied
to the simulation as an estimate of the systematic uncertainty.

Differences in both \KL and \KS production rates of data and MC are taken into account by adjusting the 
inclusive $D\to K^0 X$ and $D_s\to K^0 X$ branching fractions.  
The associated systematic uncertainty is assessed by varying these branching fractions 
within their uncertainties.

\subsection{Signal and background simulation}
\subsubsection{Signal simulation}
\label{sec:theosys}
Knowledge of the details of inclusive \Bxulnu decays is crucial to several aspects of the analysis:
the fraction of events within the selected kinematic region 
depends on the signal kinematics over the full phase space.
Specifically, the efficiencies 
$\epsilon_{u}$ and $\epsilon_\mathrm{kin}$ 
rely on accurate MC simulation, because the particle multiplicities, 
momenta, and angles depend on the hadronization model for the 
hadronic states $X_u$.

To simulate the signal \Bxulnu decays we have chosen the prescription 
by De Fazio and Neubert~\cite{De Fazio:1999sv}.
Different choices of the parameterization for the Fermi motion of the $b$ quark 
inside the \B meson (Section~\ref{sec:signalgene}) lead to different spectra of the hadron 
mass \mx\ and lepton momentum \Pl .
We estimate the impact of these choices by repeating the analysis 
with shape function parameters set to values of
\lonesf and \lbarsf\ corresponding to the  contour of the $\Delta \chi^2 =1$ error
ellipse~\cite{belleOPE}.  
To assess the impact of the choice of the SF ansatz, we repeat this procedure for a 
different SF ansatz~\cite{De Fazio:1999sv}.

Since the simulation of \Bxulnu decays is a hybrid of exclusive decays to low-mass charmless 
mesons and inclusive decays to higher-mass states $X_u$, the relative contributions of the various 
decays impact the overall kinematics and thereby the efficiencies. We evalute the impact of 
varying the branching fractions of the exclusive charmless
semileptonic \B decays by one standard deviation. 

The signal losses caused by the kaon veto depend on the production rate of kaons in these decays.
In the MC simulation, the number of $K^+$ and $K_S^0$ in the signal decays is set by the probability 
of producing \ssbar\ quark pairs from the vacuum.
The fraction of \ssbar\ events is about 12.0\% for the nonresonant component of the signal and is 
fixed by the parameter $\gamma_s$ in the fragmentation by {\sc jetset}~\cite{Sjostrand:1994yb}. 
This parameter has been measured by two experiments at center of mass energies between 12 and 
36~\gev as $\gamma_s=0.35 \pm 0.05$~\cite{Althoff:1984iz}, $\gamma_s=0.27 \pm 0.06$~\cite{Bartel:1983qp}.  
We adopt the value $\gamma_s=0.3$ and estimate
the systematic uncertainty by varying the fraction of \ssbar\ events by~$\pm30\%$.

The theoretical uncertainty  due to the lower limit on  the lepton spectrum  is largely accounted 
for by the reweighting of events for the assessment of 
the theoretical uncertainty related to the Fermi motion.

\subsubsection{Branching fractions for \B and $D$ decays}
The exclusive semileptonic branching fractions for \Bxclnu
decays and the hadronic mass spectra for these decays are crucial for
the determination of the yield of the inclusive normalization sample
and the \Bxclnu background.  Exclusive \B and $D$ branching
fractions used in the MC simulation differ slightly from the
world averages~\cite{PDG2010}; this difference is corrected by
reweighting events in the simulation. The branching fraction for the sum of semileptonic
decays to nonresonant $D^{(*)} \pi$ or broad $D^{**}$ states is taken
as the difference between the total semileptonic rate and the other
well measured branching fractions, and amounts to about 1.7\%.

Similarly, branching fractions and decay distributions for hadronic and semileptonic
$D$ meson decays affect the measurement of $\Delta R_{u/sl}$. The effect is 
different  for neutral and charged $B$ mesons, because \Bzb decays mostly into
charged $D$ mesons while \Bm decays almost always into neutral charm mesons.

Likewise, uncertainties on the form factors for \Bdbslnu decays are taken into 
account by repeating the analysis with changes of the form factor values by their 
experimental uncertainties~\cite{Aubert:2008yv}. 
For \Bdsslnu decays, the uncertainties on the form factor have not been specified. 
Thus, we perform the fits with the ISGW2~\cite{isgw2} parameterization of the form factors and take 
the difference with respect to the default fits as systematic uncertainty. 

The uncertainty related to the $\lambda_{D^{**}}$ parameter introduced in Eq.~(\ref{eq:lambda}) 
has been estimated by varying it within its uncertainty, and taking
the difference with respect to the default fits as systematic uncertainty.

\subsubsection{Combinatorial background subtraction and normalization}
For the fits to the \mes\ distributions in individual bins of a given kinematic variable, all 
parameters other than event yields and the ARGUS shape are fixed to values determined from 
distributions obtained from the full signal sample. 
To estimate the systematic uncertainty due
this choice of parameters, their values are varied within their 
statistical uncertainty, taking correlations into account. 
We estimate the effect of the combinatorial background subtraction by
determining it on a simulated sample by means of Monte Carlo truth
information, and getting the signal yields on data by subtraction.
The differences relative to the default fit are taken as systematic uncertainties.

Finally, the uncertainty on the knowledge of the total semileptonic branching fraction adds
1.4\% to the assessment of our systematic uncertainty.

In summary, the smallest statistical and systematic uncertainties are 
achieved for the $\mX<1.55$~\gevcc region, which has an acceptance that 
is reduced by 40\% 
with respect to the region defined by $\Pl > 1.0$~\gevc,
but has the best separation of signal and background. 
The dominant systematic uncertainty for samples
with no phase space restrictions, except for $\Pl > 1.0$~\gevc,
is due to the uncertainty on the shape function 
parameters which impact the differential \Q and \Pl distributions.
\section{Extraction of \Vub}
\label{sec:theonew}

\subsection{QCD corrections}
\label{sec:qcdcorrections}
We extract \Vub from the measurements of the partial branching fractions $\Delta \BR(\Bxulnu)$ 
by relying on QCD predictions.
In principle, the total rate for \Bxulnu decays can be calculated based on heavy quark 
expansions (HQE) in powers of $1/m_b$ with uncertainties at the level of 5\%, in a similar way as 
for \Bxclnu decays.
Unfortunately, the restrictions imposed on the phase space to reduce the large background from 
Cabibbo-favored decays spoil the HQE convergence. Perturbative and nonperturbative corrections are 
drastically enhanced and the rate becomes sensitive to the Fermi motion of the $b$ quark inside the 
$B$ meson, introducing terms that are not suppressed by powers of $1/m_b$. In practice,  
nonperturbative SFs are introduced.
The form of the SFs cannot be calculated from first principles.  Thus, knowledge of these SFs
relies on global fits performed by several collaborations to moments of the lepton energy and 
hadronic invariant mass in semileptonic $B$ decays, and of the photon energy in radiative 
$B\to X_s \gamma$ inclusive decays~\cite{OPE1,OPE2,OPE3}.
We adopt results of the global fits to published measurements of moments, performed in the 
kinetic renormalization scheme, specifically the $b$ quark mass $m_b^{kin} = (4.560 \pm 0.023)$~\gevcc 
and the mean value of the $b$ quark momentum operator 
$\mu_{\pi}^{2 (kin)} = (0.453 \pm 0.036)$~\gevccsq~\cite{HFAG2011,belleOPE}. 
Due to confinement and nonperturbative effects the quantitative values of the quark mass and other 
HQE parameters are specific to the theoretical framework in which it is defined. Thus the results 
of the global fits need to be translated to other schemes, depending on the QCD calculation 
used to extract \Vub. 
In the following, we determine \Vub based on four different QCD calculations. 
The numerical calculations are based on computer code kindly provided by the authors.

The measured partial branching fractions $\Delta \BR(\Bxulnu)$ are related to \Vub via the 
following equation,
\begin{equation}
\Vub  = \sqrt{\frac{\Delta \BR(\Bxulnu)}{\tau_B \; \Delta\Gamma_{\mathrm{theory}}}},\\ 
\label{eq:vub}
\end{equation}
where $\Delta\Gamma_{\mathrm{theory}}$, the theoretically predicted \Bxulnu rate for the 
selected phase space region, is
based on different QCD calculations, and the \B\ meson lifetime is $\tau_B = 1.582 \pm 0.007$~\ps~\cite{HFAG2011}.
We adopt the uncertainties 
on $\Delta\Gamma_{\mathrm{theory}}$ as assessed by the authors. 
It should be noted that the systematic uncertainty on the branching fraction that is related to the 
uncertainties on the SF parameterization are fully correlated to 
the theoretical uncertainties discussed here.

The calculated decay rates $\Delta\Gamma_{\mathrm{theory}}$ and the resulting 
\Vub values are shown for the various kinematic regions in Tables~\ref{tab:VUB} 
and~\ref{tab:deltaB}, separately for the four different QCD calculations.

\subsubsection{BLNP calculation}
The theoretical uncertainties~\cite{Lange:2005yw,Bosch:2004th,Bosch:2004cb} 
arise from the uncertainty on $m_b$, $\mu_{\pi}^2$ and other nonperturbative corrections,
the functional form of the leading and the subleading SFs, the variation of the 
matching scales, and the uncertainty on the estimated contribution from weak annihilation processes. 
The dominant contributions are due to the uncertainties on $m_b$, and $\mu_{\pi}^2$. 
These parameters need to be translated to the shape function renormalization scheme, 
for which $m_b^{(\mathrm{SF})} = (4.588 \pm 0.025)$~\gevcc and 
$\mu_{\pi}^{2 (\mathrm{SF})} = (0.189 \pm 0.051)~\mathrm{GeV^2}$. 
The stated errors include the uncertainties due to higher order terms 
which are neglected in the translation from one scheme to another. 

A recent calculation at NNLO~\cite{Greub:2009sv} indicates that the differences with respect to NLO
calculations are rather large. They would increase the value of \Vub by about 8\%, 
suggesting that the current uncertainties are underestimated. 
Similar effects might also be present for other QCD calculations, 
but estimates are not yet available.

\subsubsection{DGE calculation}
The theoretical uncertainties~\cite{Andersen:2005mj,Gardi:2008bb} 
arise from the uncertainty on $\alpha_s$, 
the uncertainty on $m_b$, and other nonperturbative corrections, for instance,
the variation of the matching scales, and the uncertainty on the weak annihilation. 
The dominant error is the uncertainty on $m_b$ for which the \MSb\ renormalization scheme is used.
Therefore the results of the global fit had to be translated to the \MSb\  scheme,  
$m_b^{\MSb} = (4.194 \pm 0.043)$~\gevcc, where the uncertainty includes the uncertainty on the translation.

\subsubsection{GGOU calculation}
The theoretical uncertainties~\cite{GGOU} in the determinations of the widths and \Vub from 
the GGOU calculations
arise from the uncertainty on $\alpha_s$, $m_b$, and $\mu_{\pi}^2$, plus various nonperturbative 
corrections: the modeling of the $q^2$ tail and choice of the scale $q_*^2$, 
the functional form of the distribution functions, and the uncertainty on the weak annihilation
rate. The dominant error originates from the uncertainties on $m_b$ and $\mu_{\pi}^2$. 
Since GGOU calculations are based on the kinetic renormalization scheme, there is no need for translation. 

\subsubsection{ADFR calculation}
The ADFR calculation~\cite{Aglietti,Aglietti:2006yb} 
relates $\Delta{\cal{B}}(\Bxulnu)$ to \Vub in a way that is
different from the other three calculations discussed above.  
In the framework of ADFR, the partial branching ratio is expressed in terms of $R_{c/u}$, 
\begin{equation}
\label{eq:ADFR_B}
\Delta \BR(\Bxulnu) = \frac{\BR(\Bxlnu)}{1+R _{c/u}}W,
\end{equation}
where $W = \Delta \Gamma(\Bxulnu) /\Gamma(\Bxulnu)$ is the fraction of the charmless branching 
fraction in a selected kinematic region and \BR(\Bxlnu) is the total semileptonic branching fraction. 
$R _{c/u}$ is related to \Vub as 
\begin{equation}
\label{eq:ADFR_R}
R_{c/u} = \frac{|V_{cb}|^{2}}{|V_{ub}|^{2}}I(\rho)G(\alpha_{s},\rho).
\end{equation}
The function $I(\rho)$ accounts for the suppression of phase space due to $m_c$ and
$I(\rho) = 1 - 8\rho + 12\rho^2\log(1/\rho) + 8\rho^2 - \rho^4, $
with $\rho \equiv m_c^2/m_b^2 \approx 0.1$. 
The factor $G(\alpha_{s},\rho)$ contains corrections suppressed by powers of $\alpha_{s}$ and powers 
of $\rho$, 
\begin{equation}
G(\alpha_{s},\rho) = 1 + \sum_{n=1}^{\infty}G_n(\rho)\alpha_{s}^n,
\end{equation}
with $G_n(0)=0$. 
The errors of the ADFR calculations arise from the uncertainty in $\alpha_s$, $\Vcb$, the quark masses  
$m_b$ and $m_c$, and the uncertainty on \BR(\Bxlnu). 
The dominant uncertainty is due to the uncertainty on the mass $m_c$. 

\subsection{\Vub extraction}
We present the results for \Vub with statistical, systematic and theoretical uncertainties in
Table~\ref{tab:VUB}. Values of \Vub extracted from partial branching fractions 
for samples with the lower limit on the lepton momentum \Pl\ varying from 1.0~\gevc to 2.4~\gevc
are tabulated in Table~\ref{tab:deltaB}. 
The different values of \Vub\ are consistent within one standard deviation
and equally consistent with the previous \babar\ measurements of 
\Vub on inclusive charmless semileptonic $B$ 
decays~\cite{babar_inclusive,Aubert:2005im,Aubert:2005mg} as well 
as a similar measurement by the Belle Collaboration~\cite{Belle_multivariate}.

Our result on the study of the lepton spectrum above 2~\gevc can be compared to
what \babar~\cite{Aubert:2005im,Aubert:2005mg}, Belle~\cite{Limosani:2005pi} 
and CLEO~\cite{Bornheim:2002du} have published on the analysis of the lepton endpoint spectrum 
in untagged $B$ decays. Experimental uncertainties are comparable, as well as theoretical uncertainties, which are
quite large in this region of phase space. The values of \Vub obtained with such different techniques 
agree very well.

We have evaluated the correlations of the measurements of \Vub in selected regions of phase space
taking into account the experimental and theoretical procedures, as presented in Table~\ref{tab:statcorr}. 
The theoretical correlations have been obtained for the BLNP calculations by taking several values of the 
heavy quark parameters within their uncertainties and computing the correlation of the acceptance for 
pairs of phase space regions. The resulting correlation coefficients are in all cases greater than 97\%. 
It is assumed that the correlations are also close to 100\% for the other three theory calculations. 

We choose to quote the \Vub value corresponding to the most inclusive measurement, 
namely the one based on the two-dimensional fit of the \mX\ -- \Q distribution with no phase space 
restrictions, except for $\Pl > 1.0$~\gevc. We calculate the arithmetic average 
of the values and uncertainties obtained with the different theoretical calculations shown above 
and find 
\begin{equation}
\Vub = (4.33 \pm 0.24 \pm 0.15) \times 10^{-3}, 
\end{equation}
where the first uncertainty is experimental and the second 
theoretical.

A calculation specifically suited for phase space regions
defined by the \mX and \Q cuts~\cite{BLL} can also be considered. This
uses as input the $b$ quark mass in the 1S scheme~\cite{1S},
$m_b^{1S} = 4.704 \pm 0.029$~\gevcc, determined by a global fit in that
scheme, similar to the one described in Section~\ref{sec:qcdcorrections}. The resulting value of \Vub
for the phase space region defined by $\mX < 1.7$~\gevcc, $\Q > 8$~\gevccsq is 
$\Vub = (4.50 \pm 0.24 \pm 0.29) \times  10^{-3}$, slightly larger but still in agreement with the other
theoretical calculations.

\begin{table*}[htb]
\begin{center}
\caption{Results for \Vub\ obtained for the four different QCD calculations. The sources of the 
quoted uncertainties are experimental
statistical, experimental systematic and theory, respectively. The theoretical \Bxlnu widths, 
${\Delta\Gamma_{\mathrm{theory}}}$ in ps$^{-1}$, for the various phase space regions examined, as 
determined from the BLNP, DGE and GGOU calculations, are also shown. The ADFR calculation uses another 
methodology (see text), therefore the values for ${\Delta\Gamma_{\mathrm{theory}}}$ have been 
obtained by inverting Eq.~(\ref{eq:vub}). The~$\Pl>1\gevc$ requirement is 
implicitly assumed in the definitions of phase space regions, unless otherwise noted.}
\vspace{0.1in}
\begin{tabular}{llcc} 
\hline\hline
QCD Calculation  &  Phase Space Region & $\Delta\Gamma_\mathrm{theory}~(\mathrm{ps}^{-1})$ & $|V_{ub}| (10^{-3})$ \\ \hline 
\rule{0pt}{9pt}     & \mX $\leq 1.55$~\gevcc    				& $39.3 \mbox{}^{+4.7}_{-4.3}$ &  $ 4.17 \pm 0.15 \pm 0.12 \mbox{}^{+0.24}_{-0.24} $    \\
     & \mX $\leq 1.70$~\gevcc    							& $46.1 \mbox{}^{+5.0}_{-4.4}$ &  $ 3.97 \pm 0.17 \pm 0.14 \mbox{}^{+0.20}_{-0.20} $    \\
     & \Pplus $\leq 0.66$~\gev 							& $38.3 \mbox{}^{+4.7}_{-4.3}$ &  $ 4.02 \pm 0.18 \pm 0.16 \mbox{}^{+0.24}_{-0.23} $    \\
     BLNP &  \mX $\leq 1.70$~\gevcc, \Q $\geq 8$~\gevccsq 	& $23.8 \mbox{}^{+3.0}_{-2.4}$ &  $ 4.25 \pm 0.19 \pm 0.13 \mbox{}^{+0.23}_{-0.25} $  \\
     & \mX\ -- \Q, $\Pl>1.0$~\gevc        					& $62.0 \mbox{}^{+6.2}_{-5.0}$ &  $ 4.28 \pm 0.15 \pm 0.18 \mbox{}^{+0.18}_{-0.20} $ \\   
     & $\Pl>1.0$~\gevc        								& $62.0 \mbox{}^{+6.2}_{-5.0}$ &  $ 4.30 \pm 0.18 \pm 0.21 \mbox{}^{+0.18}_{-0.20} $  \\ 
\rule[-1pt]{0pt}{8pt}     & $\Pl>1.3$~\gevc        				& $52.8 \mbox{}^{+5.3}_{-4.3}$ &  $4.29 \pm 0.18 \pm 0.20 \mbox{}^{+0.19}_{-0.20} $  \\ \hline

\rule{0pt}{9pt}     & \mX $\leq 1.55$~\gevcc    				& $35.3 \mbox{}^{+3.3}_{-3.5}$  & $4.40 \pm 0.16 \pm 0.12 \mbox{}^{+0.24}_{-0.19} $    \\
     & \mX $\leq 1.70 $~\gevcc    							& $42.0 \mbox{}^{+4.8}_{-4.8}$  & $4.16 \pm 0.18 \pm 0.14 \mbox{}^{+0.26}_{-0.22} $    \\
     & \Pplus $\leq 0.66$~\gev 							& $36.9 \mbox{}^{+5.5}_{-5.8}$ &  $4.10 \pm 0.19 \pm 0.17 \mbox{}^{+0.37}_{-0.28} $    \\
     DGE  & \mX $\leq 1.70$~\gevcc, \Q $\geq 8$~\gevccsq          & $24.4 \mbox{}^{+2.4}_{-2.0}$  & $4.19 \pm 0.19 \pm 0.12 \mbox{}^{+0.18}_{-0.19} $ \\ 
     &   \mX\ -- \Q, $\Pl>1.0$~\gevc    						& $58.7 \mbox{}^{+3.5}_{-3.2}$  &  $ 4.40 \pm 0.16 \pm 0.18 \mbox{}^{+0.12}_{-0.13} $    \\ 
     & $\Pl>1.0$~\gevc        								& $58.7 \mbox{}^{+3.5}_{-3.2}$  &  $ 4.42 \pm 0.19 \pm 0.23 \mbox{}^{+0.13}_{-0.13} $  \\ 
\rule[-1pt]{0pt}{8pt}     & $\Pl>1.3\gevc$       				& $50.4 \mbox{}^{+3.3}_{-3.0}$  &  $4.39 \pm 0.19 \pm 0.20 \mbox{}^{+0.15}_{-0.14} $  \\ \hline

\rule{0pt}{9pt}     & \mX $\leq 1.55$~\gev    					& $41.0 \mbox{}^{+4.6}_{-3.8}$  &  $4.08 \pm 0.15 \pm 0.11 \mbox{}^{+0.20}_{-0.21} $    \\
     & \mX $\leq 1.70$~\gev    							& $46.8 \mbox{}^{+4.2}_{-3.6}$  &  $3.94 \pm 0.17 \pm 0.14 \mbox{}^{+0.16}_{-0.17} $    \\
     & \Pplus $\leq 0.66$~\gev 							& $44.0 \mbox{}^{+8.6}_{-6.3}$  &  $3.75 \pm 0.17 \pm 0.15 \mbox{}^{+0.30}_{-0.32} $    \\
     GGOU &  \mX $\leq 1.70$~\gevcc, \Q $\geq 8$~\gevccsq 	& $24.7 \mbox{}^{+3.2}_{-2.4}$  &  $4.17 \pm 0.18 \pm 0.12 \mbox{}^{+0.22}_{-0.25} $ \\
     & \mX\ -- \Q, $\Pl>1.0$~\gevc     						& $60.2 \mbox{}^{+3.0}_{-2.5}$  &  $4.35 \pm 0.16 \pm 0.18 \mbox{}^{+0.09}_{-0.10} $   \\
     & $\Pl>1.0$~\gevc        								& $60.2 \mbox{}^{+3.0}_{-2.5}$  &  $4.36 \pm 0.19 \pm 0.23 \mbox{}^{+0.09}_{-0.10} $  \\ 
     \rule[-1pt]{0pt}{8pt}    & $\Pl>1.3$~\gevc        				& $51.8 \mbox{}^{+2.8}_{-2.3}$  &  $4.33 \pm 0.18 \pm 0.20 \mbox{}^{+0.10}_{-0.11} $  \\ \hline

     \rule{0pt}{9pt}     & \mX $\leq 1.55$~\gev    				& $47.1 \mbox{}^{+5.2}_{-4.3} $  &  $3.81 \pm 0.14 \pm 0.11 \mbox{}^{+0.18}_{-0.20} $    \\
     & \mX $\leq 1.70$~\gev    							& $52.3 \mbox{}^{+5.4}_{-4.5} $  &  $3.73 \pm 0.16 \pm 0.13 \mbox{}^{+0.17}_{-0.18} $    \\
     & \Pplus $\leq 0.66$~\gev 							& $48.9 \mbox{}^{+5.6}_{-4.6} $  &  $3.56 \pm 0.16 \pm 0.15 \mbox{}^{+0.18}_{-0.19} $    \\
     ADFR &  \mX $\leq 1.70$~\gevcc, \Q $\geq 8$~\gevccsq 	& $30.9 \mbox{}^{+3.0}_{-2.5} $  &  $3.74 \pm 0.16 \pm 0.11 \mbox{}^{+0.16}_{-0.17} $  \\
     &  \mX\ -- \Q, $\Pl>1.0$~\gevc         					& $62.0 \mbox{}^{+5.7}_{-5.0} $  &  $4.29 \pm 0.15 \pm 0.18 \mbox{}^{+0.18}_{-0.19} $   \\ 
     & $\Pl>1.0$~\gevc        								& $62.0 \mbox{}^{+5.7}_{-5.0} $  &  $4.30 \pm 0.19 \pm 0.23 \mbox{}^{+0.18}_{-0.19} $  \\ 
     \rule[-1pt]{0pt}{8pt}     & $\Pl>1.3$~\gevc        			& $53.3 \mbox{}^{+5.1}_{-4.4} $  &  $4.27 \pm 0.18 \pm 0.19 \mbox{}^{+0.18}_{-0.19} $  \\ \hline

\hline\hline
\end{tabular}
\label{tab:VUB}
\end{center}
\end{table*}

\subsection{Limits on weak annihilation}
The measurements of  $\Delta \BR(\Bxulnu)$, separately for neutral and charged \B mesons, 
are summarized in Table~\ref{tab:inputdeltaB0Bp} for the various kinematic selections.
These results are used to test isospin invariance, based on the ratio
\begin{equation}
R = \frac{\Delta \Gamma^-}{\Delta \Gamma^0}= \frac{\tau^0}{\tau^-} \; \frac{\Delta{\cal B}(\Bm\to\X_u\ell\nu)} {\Delta{\cal B}(\Bzb\to\X_u\ell\nu)},\\
\label{eq:rwa}
\end{equation}
where ${\tau^-}/{\tau^0}=1.071\pm 0.009$~\cite{PDG2010} is the ratio of the lifetimes 
for \Bm and \Bzb. 
For the $\mX < 1.55$~\gevcc selection, we obtain $R-1 = 0.03 \pm 0.15 \pm 0.18$, where 
the first uncertainty is statistical and the second is systematic. This result is 
consistent with zero; 
similar results, with larger uncertainties, are obtained for the other regions of phase space
listed in Table~\ref{tab:inputdeltaB0Bp}.
Thus, we have no evidence for a difference between partial decay rates for \Bm and \Bzb. 
If we define the possible contribution of the weak annihilation as 
$\Delta\Gamma_\mathit{WA}=\Delta \Gamma^--\Delta\Gamma^0$, 
its relative contribution to the partial decay width $\Delta \Gamma$ for $\Bxulnu$ decays 
is $\Delta{\Gamma_\mathit{WA}}/{\Delta\Gamma}=R-1$.
With $f_\mathit{WA}$ defined as the fraction of weak annihilation contribution for a specific kinematic 
region and $f_u$ defined as the fraction of $\Bxulnu$ events predicted for that region, 
we can write $\Delta\Gamma_\mathit{WA}=f_\mathit{WA}\Gamma_\mathit{WA}$
and  $\Delta\Gamma=f_u \Gamma$, where $\Gamma$ is the total decay width of \Bxulnu decays. 
Thus the relative contribution of the weak annihilation is
\begin{equation}
\frac{\Gamma_\mathit{WA}}{\Gamma}=\frac{f_u}{f_\mathit{WA}} \; (R-1).\\
\label{eq:rwag}
\end{equation}
Since the weak annihilation is expected to be confined to the high
\Q region, it is reasonable to assume $f_\mathit{WA}=1.0$ for all the kinematic selections.
We adopt the prediction for $f_u$ by De Fazio-Neubert (see Section~\ref{sec:signalgene})
and place limits on ${\Gamma_\mathit{WA}}/{\Gamma}$.
The most stringent limit is obtained for the  selection
$\mX<1.55$~\gevcc, namely $-0.17\le ({\Gamma_\mathit{WA}}/{\Gamma})<0.19$ at 90\% 
confidence level (C.L.).
This model-independent limit on WA is consistent, but weaker than the limit derived by the CLEO
collaboration~\cite{WACleo} on the basis of an assumed \Q distribution.  
Both limits are larger than the theoretical limits, estimated from $D$
and $D_s$ semileptonic decay rates, of 3\%~\cite{Bigi:1993bh,Voloshin:2001xi}, and the more recent and
stringent one of less than 2\%~\cite{Ligeti:2010vd,GambinoWA}.

\begin{sidewaystable*}[h]
\centering
\caption{Summary of the fitted number of events $N_u$, the efficiencies, the partial branching fractions $\Delta \BR(\Bxulnu)$ and \Vub\ ($10^{-3}$) based on four 
different QCD calculations of the hadronic matrix element as a function of the lower limit on the lepton momentum \Pl. 
The uncertainties on $\Delta \BR(\Bxulnu)$ are statistical and systematic, those for \Vub\ are statistical, systematic and theoretical. 
The uncertainties on all other parameters are statistical. \Vub values for BLNP and GGOU are not 
provided above 2.2~\gevc due to large uncertainties.}
\vspace{0.1in}
\begin{tabular*}{\textwidth}{c@{\extracolsep{\fill}}ccccccccc}
\hline\hline\vspace{1mm}
$p^*_{\ell_\mathrm{min}}$ &   &  &    & \raisebox{-.2ex}[0cm][0cm]{$\Delta \BR(\Bxulnu)$} & \Vub BLNP & \Vub GGOU &\Vub DGE & \Vub ADFR\\
\raisebox{.2ex}[0cm][0cm]{(\gevc)} & \raisebox{1.5ex}[0cm][0cm]{$N_u$} & \raisebox{1.5ex}[0cm][0cm]{$\epsilon_\mathrm{sel}^u \epsilon_\mathrm{kin}^u$} & \raisebox{1.5ex}[0cm][0cm]{$\frac{\epsilon_\ell^\mathrm{sl} \epsilon_\mathrm{tag}^\mathrm{sl}} {\epsilon_\ell^u \epsilon_\mathrm{tag}^u }$} & \raisebox{.2ex}[0cm][0cm]{($10^{-3}$)} & \raisebox{.2ex}[0cm][0cm]{($10^{-3}$)} & \raisebox{.2ex}[0cm][0cm]{($10^{-3}$)} & \raisebox{.2ex}[0cm][0cm]{($10^{-3}$)} & \raisebox{.2ex}[0cm][0cm]{($10^{-3}$)} \\\hline \vspace{0.03in}
\rule{0pt}{9pt}{\bf 1.0} & $1470 \pm 130$	& $0.342 \pm 0.002$ & $1.18 \pm 0.03$ & 
$1.81 \pm 0.16 \pm 0.19$ & $4.30 \pm 0.18 \pm 0.21 \mbox{}^{+0.18}_{-0.20} $  &  $4.36 \pm 0.19 \pm 0.23 \mbox{}^{+0.09}_{-0.10} $  &  $4.42 \pm 0.19 \pm 0.23 \mbox{}^{+0.13}_{-0.13} $  &  $4.30 \pm 0.19 \pm 0.23 \mbox{}^{+0.18}_{-0.19} $ \\ \vspace{0.03in}
{\bf 1.1} & $1440 \pm 127$	& $0.345 \pm 0.002$ & $1.18 \pm 0.19$ &
$1.75 \pm 0.15 \pm 0.18$ & $4.32 \pm 0.17 \pm 0.19 \mbox{}^{+0.18}_{-0.20} $  &  $4.38 \pm 0.19 \pm 0.22 \mbox{}^{+0.09}_{-0.11} $  &  $4.44 \pm 0.19 \pm 0.23 \mbox{}^{+0.14}_{-0.13} $  &  $4.34 \pm 0.19 \pm 0.22 \mbox{}^{+0.20}_{-0.20} $ \\ \vspace{0.03in}
{\bf 1.2} & $1421 \pm 124$	& $0.353 \pm 0.002$ & $1.18 \pm 0.05$ & 
$1.69 \pm 0.14 \pm 0.18$ & $4.36 \pm 0.17 \pm 0.22 \mbox{}^{+0.19}_{-0.21} $  &  $4.41 \pm 0.18 \pm 0.23 \mbox{}^{+0.10}_{-0.11} $  &  $4.47 \pm 0.19 \pm 0.24 \mbox{}^{+0.14}_{-0.14} $  &  $4.36 \pm 0.18 \pm 0.23 \mbox{}^{+0.20}_{-0.20} $ \\ \vspace{0.03in}
{\bf 1.3} & $1329 \pm 121$	& $0.363 \pm 0.002$ & $1.18 \pm 0.09$ & 
$1.53 \pm 0.13 \pm 0.14$ & $4.29 \pm 0.18 \pm 0.20 \mbox{}^{+0.19}_{-0.20} $  &  $4.33 \pm 0.18 \pm 0.20 \mbox{}^{+0.10}_{-0.11} $  &  $4.39 \pm 0.19 \pm 0.20 \mbox{}^{+0.15}_{-0.14} $  &  $4.27 \pm 0.18 \pm 0.19 \mbox{}^{+0.18}_{-0.19} $ \\ \vspace{0.03in}
{\bf 1.4} & $1381 \pm 114$	& $0.368 \pm 0.002$ & $1.18 \pm 0.04$ & 
$1.58 \pm 0.13 \pm 0.14$ & $4.52 \pm 0.17 \pm 0.18 \mbox{}^{+0.20}_{-0.22} $  &  $4.55 \pm 0.19 \pm 0.20 \mbox{}^{+0.11}_{-0.12} $  &  $4.61 \pm 0.19 \pm 0.20 \mbox{}^{+0.16}_{-0.15} $  &  $4.48 \pm 0.18 \pm 0.20 \mbox{}^{+0.21}_{-0.21} $ \\ \vspace{0.03in}
{\bf 1.5} & $1383 \pm 107$	& $0.378 \pm 0.003$ & $1.19 \pm 0.02$ & 
$1.53 \pm 0.12 \pm 0.14$ & $4.66 \pm 0.16 \pm 0.18 \mbox{}^{+0.21}_{-0.23} $  &  $4.67 \pm 0.18 \pm 0.21 \mbox{}^{+0.11}_{-0.14} $  &  $4.74 \pm 0.19 \pm 0.22 \mbox{}^{+0.17}_{-0.17} $  &  $4.59 \pm 0.18 \pm 0.21 \mbox{}^{+0.21}_{-0.22} $ \\ \vspace{0.03in}
{\bf 1.6} & $1248 \pm   99$	& $0.390 \pm 0.003$ & $1.17 \pm 0.03$ & 
$1.35 \pm 0.10 \pm 0.13$ & $4.64 \pm 0.17 \pm 0.20 \mbox{}^{+0.21}_{-0.23} $  &  $4.63 \pm 0.17 \pm 0.22 \mbox{}^{+0.12}_{-0.15} $  &  $4.69 \pm 0.17 \pm 0.23 \mbox{}^{+0.18}_{-0.18} $  &  $4.52 \pm 0.17 \pm 0.22 \mbox{}^{+0.21}_{-0.21} $ \\ \vspace{0.03in}
{\bf 1.7} & $1158 \pm  90$		& $0.404 \pm 0.003$ & $1.16 \pm 0.03$ & 
$1.22 \pm 0.09 \pm 0.12$ & $4.71 \pm 0.17 \pm 0.20 \mbox{}^{+0.22}_{-0.24} $  &  $4.68 \pm 0.17 \pm 0.23 \mbox{}^{+0.14}_{-0.16} $  &  $4.73 \pm 0.17 \pm 0.23 \mbox{}^{+0.21}_{-0.19} $  &  $4.53 \pm 0.17 \pm 0.22 \mbox{}^{+0.21}_{-0.22} $ \\ \vspace{0.03in}
{\bf 1.8} & $1043 \pm  80$		& $0.418 \pm 0.003$ & $1.16 \pm 0.04$ & 
$1.07 \pm 0.08 \pm 0.10$ & $4.79 \pm 0.17 \pm 0.21 \mbox{}^{+0.23}_{-0.25} $  &  $4.71 \pm 0.18 \pm 0.22 \mbox{}^{+0.15}_{-0.18} $  &  $4.75 \pm 0.18 \pm 0.22 \mbox{}^{+0.23}_{-0.20} $  &  $4.51 \pm 0.17 \pm 0.21 \mbox{}^{+0.21}_{-0.22} $ \\ \vspace{0.03in}
{\bf 1.9} & $ 845 \pm  69$		& $0.430 \pm 0.004$ & $1.14 \pm 0.06$ & 
$0.85 \pm 0.07 \pm 0.10$ & $4.76 \pm 0.18 \pm 0.23 \mbox{}^{+0.23}_{-0.26} $  &  $4.63 \pm 0.19 \pm 0.27 \mbox{}^{+0.17}_{-0.21} $  &  $4.64 \pm 0.19 \pm 0.27 \mbox{}^{+0.26}_{-0.22} $  &  $4.36 \pm 0.18 \pm 0.26 \mbox{}^{+0.20}_{-0.21} $ \\ \vspace{0.03in}
{\bf 2.0} & $ 567 \pm  56$		& $0.457 \pm 0.004$ & $1.11 \pm 0.04$ &
$0.55 \pm 0.05 \pm 0.06$ & $4.41 \pm 0.20 \pm 0.20 \mbox{}^{+0.24}_{-0.28} $  &  $4.22 \pm 0.19 \pm 0.23 \mbox{}^{+0.18}_{-0.23} $  &  $4.19 \pm 0.19 \pm 0.23 \mbox{}^{+0.27}_{-0.23} $  &  $3.86 \pm 0.18 \pm 0.21 \mbox{}^{+0.19}_{-0.20} $ \\ \vspace{0.03in}
{\bf 2.1} & $ 432 \pm  44$		& $0.474 \pm 0.005$ & $1.07 \pm 0.03$ & 
$0.42 \pm 0.04 \pm 0.05$ & $4.68 \pm 0.22 \pm 0.24 \mbox{}^{+0.31}_{-0.37} $  &  $4.37 \pm 0.21 \pm 0.26 \mbox{}^{+0.24}_{-0.32} $  &  $4.25 \pm 0.20 \pm 0.25 \mbox{}^{+0.35}_{-0.29} $  &  $3.82 \pm 0.18 \pm 0.23 \mbox{}^{+0.18}_{-0.19} $ \\ \vspace{0.03in}
{\bf 2.2} & $ 339 \pm  29$		& $0.499 \pm 0.007$ & $1.02 \pm 0.04$ & 
$0.33 \pm 0.03 \pm 0.03$ & $5.51 \pm 0.22 \pm 0.23 \mbox{}^{+0.57}_{-0.68} $  &  $5.02 \pm 0.23 \pm 0.23 \mbox{}^{+0.48}_{-0.61} $  &  $4.62 \pm 0.21 \pm 0.21 \mbox{}^{+0.50}_{-0.43} $  &  $4.00 \pm 0.18 \pm 0.18 \mbox{}^{+0.19}_{-0.21} $ \\ \vspace{0.03in}
{\bf 2.3} & $ 227 \pm  19$		& $0.521 \pm 0.009$ & $1.00 \pm 0.04$ & 
$0.22 \pm 0.02 \pm 0.02$ & -- 	     									    &  --									        &  $5.15 \pm 0.24 \pm 0.24 \mbox{}^{+0.92}_{-0.79} $  &  $4.17 \pm 0.19 \pm 0.19 \mbox{}^{+0.22}_{-0.25} $ \\ \vspace{0.03in}
{\bf 2.4} & $  82 \pm   9$		& $0.539 \pm 0.013$ & $1.00 \pm 0.08$ & 
$0.08 \pm 0.01 \pm 0.01$ & -- 	     									    &  --									        &  $5.11 \pm 0.34 \pm 0.34 \mbox{}^{+2.08}_{-1.92} $  &  $3.67 \pm 0.24 \pm 0.24 \mbox{}^{+0.23}_{-0.26} $ \\
\hline \hline
\end{tabular*}
\label{tab:deltaB}
\end{sidewaystable*}

\section{Conclusions}
\label{sec:conclusions}
In summary, we have measured the branching fractions for inclusive 
charmless semileptonic \B decays \Bxulnu, in various overlapping regions of 
phase space, based on the full \babar\ data sample. 
The results are presented for the full sample, and also separately for charged and neutral \B mesons.

We have extracted the magnitude of the CKM element \Vub based on
several theoretical calculations. 
Measurements in different phase space regions are consistent for all sets of
calculations, within their uncertainties.
Correlations between \Vub measurements, including both experimental and theoretical 
uncertainties are presented. They are close to 100\% for the theoretical input.

We have obtained the most precise results from the analysis based on the two-dimensional fit to 
\mX -- \Q, with no restriction other than $\Pl>1.0$~\gevc. 
The total uncertainty is about 7\%,
comparable in precision to the result recently presented 
by the Belle Collaboration~\cite{Belle_multivariate} which uses a 
multivariate discriminant to reduce the combinatorial background.
The results presented here supersede earlier \babar\ measurements based on 
a smaller tagged sample of events~\cite{babar_inclusive}.

We have found no evidence for isospin violation; the difference between the partial 
branching fractions for \Bzb and \Bm is consistent with zero. 
Based on this measurement, we place a limit on a potential contribution 
from Weak Annihilation of 19\% of the total charmless semileptonic branching faction at 90\% C.L., 
which is still larger than recent theoretical expectations~\cite{Ligeti:2010vd,GambinoWA}. 

Improvements in these measurements will require larger tagged data samples recorded with 
improved detectors and much improved understanding of the simulation of semileptonic \B decays,  
both background decays involving charm mesons as well as exclusive and inclusive decays 
contributing to the signal. Reductions in the theoretical uncertainties are expected to come
from improved QCD calculations for $\bbar \to u \ell \bar{\nu}$ and \btosgam\ transitions, 
combined with improved information on the  $b$ quark mass and measurements of radiative \B decays.

\section{Acknowledgements}
We thank Matthias Neubert, Gil Paz, Einan Gardi, Paolo Gambino, Paolo Giordano, Ugo Aglietti,
Giancarlo Ferrera, and Giulia Ricciardi for useful discussions and for providing the software tools 
and code which enabled us to compute \Vub values from measured branching fractions. 

We are grateful for the 
extraordinary contributions of our \pep2\ colleagues in
achieving the excellent luminosity and machine conditions
that have made this work possible.
The success of this project also relies critically on the 
expertise and dedication of the computing organizations that 
support \babar.
The collaborating institutions wish to thank 
SLAC for its support and the kind hospitality extended to them. 
This work is supported by the
US Department of Energy
and National Science Foundation, the
Natural Sciences and Engineering Research Council (Canada),
the Commissariat \`a l'Energie Atomique and
Institut National de Physique Nucl\'eaire et de Physique des Particules
(France), the
Bundesministerium f\"ur Bildung und Forschung and
Deutsche Forschungsgemeinschaft
(Germany), the
Istituto Nazionale di Fisica Nucleare (Italy),
the Foundation for Fundamental Research on Matter (The Netherlands),
the Research Council of Norway, the
Ministry of Education and Science of the Russian Federation, 
Ministerio de Ciencia e Innovaci\'on (Spain), and the
Science and Technology Facilities Council (United Kingdom).
Individuals have received support from 
the Marie-Curie IEF program (European Union), the A. P. Sloan Foundation (USA) 
and the Binational Science Foundation (USA-Israel).


\begin{thebibliography}{99}
\bibitem{footnote}
Charge-conjugate modes are implied throughout this paper, unless 
explicitly stated.
\end{thebibliography}
\end{document}